\documentstyle[11pt]{article}

\renewcommand{\theequation}{\arabic{section}.\arabic{equation}}

\newcommand{\half}{{\scriptstyle{\frac{1}{2}}}}
\newcommand{\LP}{\lambda \Phi^4}
\newcommand{\dl}{\delta^{(3)}({\bf r})}
\newcommand{\del}{{\mbox{\boldmath $\nabla$}}}
\newcommand{\BE}{\begin{equation}}
\newcommand{\EE}{\end{equation}}
\newcommand{\BA}{\begin{eqnarray}}
\newcommand{\EA}{\end{eqnarray}}
\newcommand{\vol}{{\sf V}}
\newcommand{\num}{{\sf N}}
\begin{document}
\begin{titlepage}
\begin{flushright}
{ \small DE-FG05-92ER40717-46 }
\end{flushright}
\vspace*{15mm}
\begin{center}
            {\LARGE{\bf $\lambda\Phi^4$ Theory from a particle-gas viewpoint}}

\vspace*{11mm}
{\Large  M. Consoli}
\vspace*{3mm}\\
{\large
Istituto Nazionale di Fisica Nucleare, Sezione di Catania \\
Corso Italia 57, 95129 Catania, Italy}
\vspace*{4mm}\\
and \\
\vspace*{4mm}
{\Large P. M. Stevenson}
\vspace*{3mm}\\
{\large T. W. Bonner Laboratory, Physics Department \\
Rice University, P.O. Box 1892, Houston, TX 77251-1892, USA}
\vspace{13mm}\\
{\bf Abstract:}
\end{center}

      We discuss the physics of the $3+1$~dimensional $\LP$ quantum field 
theory in terms of the statistical mechanics of a gas of particles (`atoms') 
that interact via a $-1/r^3$-plus-hard-core potential.  The hard-core 
potential, $\dl$, arises from the bare vertex diagram, while the attractive, 
long-range $-1/r^3$ potential is due to exchange of a particle pair via the 
$t,u$-channel ``fish'' diagram.  (Higher-order diagrams preserve this form 
of the interparticle potential.)  For sufficiently small atom mass, the 
lowest-energy state is not the `empty' state with no atoms, but a state 
with a non-zero density of spontaneously created atoms, Bose-condensed 
in the zero-momentum mode.  This corresponds to the 
spontaneous-symmetry-breaking phase transition, and the `phonon' excitations 
of the Bose condensate correspond to Higgs particles.  The important point 
is that the phase transition happens while the atom's physical mass $m$ 
is still positive: it does not wait until $m^2$ passes through zero and 
becomes negative, contrary to the assumption of a second-order transition, 
on which renormalization-group-improved perturbation theory is based.  

\vspace*{5mm}
\begin{center}
{\small e-mail: \hspace{2mm} consoli@ct.infn.it \quad 
stevenson@physics.rice.edu}
\end{center}

\end{titlepage}
 
\setcounter{page}{1}

\setcounter{equation}{0}
\section{Outline} 

     In this paper we consider the physics of $(\LP)_{3+1}$ theory 
from a ``particle'' viewpoint -- in the same sense that quantum 
electrodynamics (QED) can be viewed, not as an abstract field theory, but 
as a theory of charged particles interacting via a $1/r$ potential.  
Our aim is to gain some really `earthy' intuition about the physics.  
We begin by outlining the whole story, the component parts of which 
will be discussed in the remainder of the paper.  

     It is well known that the non-relativistic (NR) limit of $\LP$ theory 
corresponds to spinless particles that interact via a $\dl$ 
potential \cite{beg}.  Our key observation is that a qualitatively different 
potential arises from the exchange of a particle-antiparticle pair via the 
``fish'' diagram.  For massless exchanged particles this gives rise to an 
attractive, long-range, $-1/r^3$ potential.  (Including the mass yields an 
additional factor of $2mr K_1(2mr)$ which supplies an $\exp(-2mr)$ factor 
at large $r$.)  Thus, the net interparticle potential has a 
``$-1/r^3$-plus-hard-core'' form.  The hard core, a regularized form 
of the $\dl$ term, has large positive potential for $r < r_0$, where 
$1/r_0$ acts as an ultraviolet cutoff. 

     We therefore consider the statistical mechanics of a dilute gas of $\num$ 
such bosons in a box of volume $\vol$.  Without the $-1/r^3$ term this would 
be the classic hard-sphere Bose-gas problem \cite{lhy,huang,aaa}.  There is 
interesting physics -- Bose-Einstein condensation, corresponding to SSB from 
a symmetric phase with `atoms' to a broken phase where `phonons' are the 
excitations.  The spectrum of the phonons differs from that of the atoms.  
With NR kinematics, where the atom spectrum is $k^2/2m$, the phonons have 
a Bogoliubov spectrum $(k/2m)\sqrt{k^2 + 16 \pi n a}$, where $a$ is the 
atoms' scattering length and $n$ is their number density, $\num/\vol$.  
Modifying the analysis to allow for relativistic kinematics yields a phonon 
spectrum that, for $m$ small, is of normal relativistic form 
$E(k) = \sqrt{k^2 + 8 \pi n a}$.  The parameter 
\BE 
M(n) = \sqrt{8 \pi n a}
\EE
is then the phonon mass, and we can identify the `phonons' as Higgs 
particles.  The non-trivial change of spectrum survives even in the limit 
$a \to 0$, provided that $n a$ remains finite \cite{csz}.  Although this 
requires $n \to \infty$, the gas is still {\it dilute} because $n a^3$ 
vanishes; i.e., the spacing of the atoms is vastly greater than their 
scattering length.  In such a limit the approximations $n a^3 \ll 1$ 
and $k a \ll 1$ required by the analysis become exact.  
One then has a ``trivial''-but-not-entirely trivial theory \cite{csz}: 
the phonons are exactly non-interacting, but are non-trivial coherent 
states of indefinitely many atoms, and their spectrum is manifestly not 
the original atom spectrum.  

      All of this assumes, however, that the system is constrained to have 
$\num$ particles (with the interesting physics requiring $n$ large 
$\sim 1/a$).  
Relativistically, though, the `atoms' can mutually annihilate, so the 
lowest-energy state would just be an empty box.  

    However, inclusion of the $-1/r^3$ long-range attraction changes the 
situation.  It provides a potential-energy gain that could offset the 
rest-mass cost associated with $\num$ particles and the energy cost of their 
mutual hard-core replusion.  In that case, an empty box could become unstable 
to spontaneous particle creation, leading to a new vacuum state with a 
non-zero density of `atoms'.  We find that, indeed, this phenomenon 
{\it does} occur once the atom's mass $m$ is sufficiently small.  

    The attractive potential gives a negative energy contribution 
$\half \num^2 \bar{u}$, where $\bar{u}$ is the average potential energy of 
a pair.  For a $-1/r^3$ potential $\bar{u}$ is logarithmically divergent, since 
one basically finds $\bar{u} \sim \int \! d^3r (-1/r^3)/\vol$.  
Obviously, the atomic mass $m$ would cut off the integral at $1/(2m)$, but 
if we make $m$ smaller and smaller then eventually a more important 
consideration is that the long-distance interaction between two particles 
will be ``screened" by intervening particles.  The effective $r_{\rm max}$ 
will then involve the density $n = \num/\vol$.  Specifically, we argue that the 
two virtual particles being exchanged have to propagate through the 
background density of atoms and multiple collisions effectively convert 
their propagators from `atom' to `phonon' propagators.  Thus, $r_{\rm max}$ 
is not $1/(2m)$ but $1/(2 M(n))$, once we are in a regime where $M(n)$ 
is much greater than $m$.  

      The energy of the gas at zero temperature (when almost all particles 
have zero 3-momentum) then consists of three terms (i) rest masses, 
(ii) repulsion energy (obtained in the hard-sphere Bose-gas analysis 
\cite{lhy,huang,aaa}), and (iii) energy gain from attraction:  
\BE
E = \num m + c_1 \frac{\num n a}{m} + \half \num^2 \bar{u},
\EE
where $c_1$ is a numerical coefficient and $\bar{u}$ is $1/\vol$ times the 
strength of the $-1/r^3$ potential (proportional to $- \lambda^2/m^2$), 
times $\ln(r_{\rm max}^2/r_0^2) \sim - \ln(n a r_0^2)$.  Dividing by the 
volume, this gives an energy density for a state of density $n$ of the form 
\BE
{\cal E} \equiv E/\vol = n m + c_1 \frac{n^2 a}{m} + 
c_2 \frac{\lambda^2}{m^2} n^2 \ln (n a r_0^2),
\EE
where $c_2$ is another numerical coefficient.  The absence of $n^3$ 
or higher terms reflects the assumption that the gas is dilute 
($na^3 \ll 1$), rendering 3-body interactions negligible.  Indeed, 
the last term, resulting from pairwise attractive interactions, would 
have been a simple $n^2$ term (absorbable into a re-definition of $a$) 
if the long-range potential had fallen off faster than $-1/r^3$.  As 
it is, the incipient divergence results in the $\ln n$ factor.

     Upon minimizing ${\cal E}$ with respect to $n$, we see that, once 
$m$ is small enough, a state with non-zero density, $n_v$, will be 
energetically favoured over the `empty' vacuum, $n=0$.  The vital 
point is that the phase transition is first-order; it happens while $m^2$ 
is still positive, and involves a discontinuous change in the density from 
0 to $n_v$.  The usual `renormalization-group-improved perturbation theory' 
picture assumes a second-order phase transition occurring only when $m^2$ 
(the renormalized mass-squared of the symmetric vacuum) reaches zero; i.e., 
when the $n=0$ symmetric vacuum has tachyonic excitations, rendering it 
locally unstable.  

     The above description has a direct translation into quantum-field-theory 
(QFT) language; the density $n$ corresponds to $\half m \phi^2$, where 
$\phi$ is the expectation value of the field, and the scattering length 
$a$ is related to the coupling strength by $a = \lambda/(8 \pi m)$.  (Hence, 
$M^2(n) = 8 \pi a n$ translates to $\half \lambda \phi^2$, as expected.)  
The energy density as a function of density translates directly into the 
effective potential of QFT.  It consists of a $\half m^2 \phi^2$ term, 
a $\lambda \phi^4$ term, and a 
$\lambda^2 \phi^4 \ln(\half \lambda \phi^2 / \Lambda^2)$ term, where 
the ultraviolet cutoff $\Lambda$ is proportional to $1/r_0$.  

     The effective potential thus has the same form as the famous 
``one-loop'' result \cite{cw}.  Conventional wisdom holds that in $\LP$ 
theory the one-loop result is not reliable --- and, indeed, from a 
loop-expansion perspective it cannot be justified.  Nevertheless, as we have 
argued previously \cite{csz},\cite{cs}-\cite{pre}, this form is actually 
effectively exact.  The particle-gas picture clarifies the reasons for 
this exactness and provides new physical insight.  

     In order for the Higgs mass $M_h \equiv M(n\!=\!n_v)$ to 
be finite we need $\lambda$ to tend to zero as $1/\ln (\Lambda/M_h)$ 
\cite{csz},\cite{cs}-\cite{pre}.  For the phase transition to occur, the 
mass term $n m = \half m^2 \phi^2$ must not dominate the other terms; 
this requires $m^2 < \lambda M_h^2$; i.e., $m$ is infinitesimally small 
in the physical units defined by $M_h$.  It follows that $a$ will tend 
to zero, giving ``triviality.''  The new ground state is still dilute 
($n_v a$ is finite, but $a \to 0$, so $n_v a^3 \to 0$), and so the 
approximations involved all become {\it exact}.  Moreover, since $\lambda$ 
is of order $1/\! \ln \Lambda$, higher-order diagrams cannot materially 
alter the ``$-1/r^3$-plus-hard-core'' form of the interparticle potential 
(see Sect. 9).  To incorporate all-orders effects we simply let `$\lambda$' 
be the {\it effective} strength of the short-range repulsion (such that 
$a=\lambda/(8 \pi m)$ is the {\it actual}, not the Born-approximation, 
scattering length).  

      The ``particle-gas'' approach gives a clue to what goes wrong with 
the `renormalization-group-improved perturbation theory' approach.  
That approach assumes that loop diagrams basically 
renormalize the strength of the $\dl$ vertex, turning the bare coupling 
into a running coupling.  However, that is only part of the story; 
the $t,u$-channel ``fish'' diagram produces qualitatively different 
physics; long-range attraction rather than short-range repulsion.  
Higher orders then serve to renormalize both those interactions.  It is 
as though there were two operators in the game, not one.  

\setcounter{equation}{0}
\section{Preliminaries}

     We now turn to a detailed exposition of the picture just outlined.  
We shall consider the single-component $\LP$ theory, in which there is 
only a discrete symmetry $\Phi \to - \Phi$, and the particle is its own 
antiparticle.  (Generalization to the O$(N)$ case should be straightforward.)  
The Hamiltonian is: 
\BE 
\label{hamiltonian}
H = \, : \! \int \! d^3 x \left[ \frac{1}{2} \left( \dot{\Phi}^2 + 
(\del \Phi)^2 + m^2 \Phi^2 \right) + 
\frac{\lambda}{4!} \Phi^4 \right] \! : \; , 
\EE
where :$ \ldots \,$: indicates normal ordering so that the parameter $m$ 
is the {\it physical} mass of the `atoms.'  We use the term `atoms' as 
shorthand for ``particles of the symmetric phase'' (i.e., elementary 
excitations above the $\langle \Phi \rangle =0$ vacuum).  

    The aim of this paper is to gain physical insight into results 
previously obtained using field-theory formalism.  What we are after is 
not a better calculation, but some visceral physical intuition.  For 
calculational purposes the ``field language'' is preferable, but for 
intuitive insight the ``particle language'' is uniquely valuable.  We 
basically seek the ``translation'' between these two different languages.  
The ``particle language'' is obviously natural for describing the NR limit 
of the theory, and in this sense we are following the lead of B\'eg and 
Furlong, Huang, and Jackiw \cite{beg}.  However, we wish to use the particle 
language {\it without} necessarily invoking the NR limit.  As always, this 
is somewhat problematic, since ``relativistic quantum mechanics'' is not 
a completely well-defined theoretical framework.  We shall not worry overmuch 
about numerical coefficients, since it is the general {\it form} of the 
result that is the key point.  

     To illustrate the above remarks, we briefly mention a precedent 
in QED.  Consider the electron-nucleus bound-state problem: the NR limit 
is simple in ``particle'' language --- one just needs to solve the 
Schr\"odinger equation with a $-1/r$ potential.  To incorporate relativistic 
kinematics and spin effects one can replace the Schr\"odinger equation with 
the Dirac equation for a $-1/r$ potential.  However, that does not account 
for all relativistic effects; it omits effects due to virtual 
particle-antiparticle pairs, which give rise to the Lamb shift.  However, 
even those effects can be patched into the particle language; one first 
calculates how virtual pairs modify the $-1/r$ potential at short distances, 
and then considers the additional potential term as a perturbation.  Such 
a particle-language treatment, originally due to Bethe \cite{bethe}, 
captures the essential physics, though it is hard to make it into a fully 
satisfactory calculation (see \cite{bjd}, Sect. 8.7).  While a field-language 
approach is preferable for achieving a systematic calculation, the 
particle-language approach remains precious for the insight it provides.  

      The plan of the paper is as follows:  Sect. 3 discusses the 
interparticle potential; Sect. 4 reviews the hard-sphere Bose gas 
and its ``relativization''; and Sect. 5 considers the effect of including 
a $-1/r^3$ potential.  These ingredients are combined to obtain the 
energy-density expression in Sect. 6.  This is not meant to be a proper 
calculation; rather, it is a collage, cut-and-pasted from well-established 
sources, intended to reveal the essential physics.  We shamelessly ignore 
complications that only affect numerical factors, such as identical-particle 
factors of 2.  

      Sect. 7 discusses the `translations' relating the scattering 
length $a$ and density $n$ to the QFT quantities $\lambda$ and $\phi^2$.  
The translated results then yield the QFT effective potential.  Sect. 8 
examines the phase transition, and the continuum limit (how $m$ and 
$\lambda$ must scale in the cutoff $\to \infty$ limit, and what this means 
for $n$ and $a$).  Sect. 9 shows that higher-order diagrams do not 
materially alter the interparticle potential.  Our conclusions and 
outlook are summarized in Sect. 10.  

      Appendices A and B review quantum-mechanical (QM) scattering from 
a short-range potential, and from a $-1/r^3$-plus-hard-core potential, 
respectively.  Appendix C discusses field re-scaling in particle-gas 
language.  Appendix D offers an alternative version of the calculation 
which, if less physically insightful, is tidier and has the numerical 
factors properly in place.

\setcounter{equation}{0}
\section{The equivalent interparticle potential}

     The equivalence between photon exchange and the $1/r$ Coulomb potential 
(and between pion exchange and the Yukawa potential ${\rm e}^{-mr} \! /r$) 
is well known.  Quite generally, for a given QFT matrix element ${\cal M}$ 
for $2 \to 2$ scattering, we can ask ``What interparticle potential 
$V({\bf r})$, when used in QM scattering theory, would yield the same 
scattering amplitude?''  A sophisticated discussion of this question is 
given in the review of Feinberg {\it et al} \cite{feinberg}.  For our 
purposes a rather simple-minded version will suffice.  Consider the 
elastic scattering of equal mass particles, ${\bf p}_1, {\bf p}_2 
\to {\bf p}_1', {\bf p}_2'$, in the centre-of-mass frame 
(${\bf p}_1 = - {\bf p}_2 \equiv {\bf p}$).  Define 
$E = \sqrt{{\bf p}^2 + m^2}$ and let ${\bf q}$ be the momentum transfer 
${\bf p}_1' - {\bf p}_1$, and the scattering angle be $\theta$.
Let us compare the QFT formula for the differential cross section 
\cite{fnteid}:
\BE
\label{dsdoqft}
     \frac{d \sigma}{d \Omega} = \frac{1}{256 \pi^2} \frac{1}{E^2} 
\, \mid \! {\cal M} \! \mid^2,
\EE
with the corresponding formula in QM:
\BE
\label{dsdoqm}
     \frac{d \sigma}{d \Omega} = \, \mid \! f(\theta) \! \mid^2,
\EE
where $f(\theta)$, the scattering amplitude, is given, in Born approximation, 
by
\BE
\label{fthv}
f(\theta) = - \frac{E/2}{2 \pi} \int d^3 r \, {\rm e}^{-i {\bf q}.{\bf r}}
\, V({\bf r}).
\EE
Note that we have ``relativized'' the familiar NR QM result by replacing 
the reduced mass $\mu \equiv m/2$ by $E/2$.  This corresponds to replacing 
the Schr\"odinger equation with the ``relativized'' version 
$\hat{E}_1 + \hat{E}_2 + V({\bf r})$, with 
$\hat{E}_i \equiv \sqrt{\hat{{\bf p}}_i^2 + m^2}$, where $\hat{{\bf p}}_i$ 
is particle $i$'s momentum operator \cite{feinberg,fnterel}.  

     Comparing the two formulas (and fixing the phase factor to get a 
positive result in (\ref{vbare}) below) gives 
\BE
\int \! d^3 r \, {\rm e}^{-i {\bf q}.{\bf r}} \, V({\bf r}) =  
\frac{1}{4 E^2} {\cal M}.
\EE
Taking the 3-dimensional Fourier transform of each side gives 
\BE
\label{veq}
V({\bf r}) = \frac{1}{4 E^2} \int \! \frac{d^3 q}{(2 \pi)^3} \,
{\rm e}^{i {\bf q}.{\bf r}} \, {\cal M}.
\EE  
Note that this ``equivalent interparticle potential'' is energy dependent, 
so it is an ``effective'' concept only.  However, this caveat becomes 
unimportant in the NR regime, $E \sim m$.  Following Feinberg {\it et al} 
\cite{feinberg} we regard the $1/(4E^2)$ factor as independent of the momentum 
transfer ${\bf q}$.  Hence, $V({\bf r})$ is just a function of the relative 
postion ${\bf r}$ (conjugate to ${\bf q}$) that also depends, parametrically, 
on $E$.

     For $\lambda \Phi^4$ theory the bare vertex diagram gives 
${\cal M} = \lambda$ and the resulting interparticle potential is 
\BE
\label{vbare}
V({\bf r}) = \frac{1}{4 E^2} \lambda \dl. 
\EE

     We now ask what interparticle potential is produced by the $t$-channel 
``fish'' diagram.  First we note that if ${\cal M}$ depends only on 
$q \equiv \mid \! {\bf q} \! \mid $ then the angular integrations in 
(\ref{veq}) yield 
\BE
V(r) = \frac{1}{4 E^2} \frac{1}{(2 \pi)^2} \int_0^\infty q^2 dq 
\frac{2 \sin qr}{q r} {\cal M}(q)
\EE
\BE 
\label{veq2}
\quad \quad \quad = \frac{1}{8 \pi^2 E^2} \, \frac{1}{r^3} 
\int_0^\infty dy \, y \sin y \, {\cal M}(q=y/r).
\EE
Note that $V(r)$ is spherically symmetric and naturally has a $1/r^3$ 
factor.  

    The matrix element for the $t$-channel ``fish'' diagram, neglecting 
the masses of the exchanged particles for now, is just 
\BE
\label{mtexch}
{\cal M}_{t-{\rm exch}}(q) = \frac{\lambda^2}{16 \pi^2} \ln(q/\Lambda),
\EE
where $\Lambda$ is an ultraviolet cutoff.  Substituting in Eq. (\ref{veq2}) 
yields an integral that is not properly convergent, but which can be made 
well defined by including a convergence factor ${\rm e}^{-\epsilon y}$ 
and then taking the limit $\epsilon \to 0$.  In this sense we have 
\BE 
\label{gr1}
\int _0^\infty dy y \sin y =0, 
\EE 
and 
\BE 
\label{gr2}
\int _0^\infty dy y \ln y \sin y = - \frac{\pi}{2}.
\EE 
(We made use of formulas 3.944.5 and 4.441.1 from Gradshteyn and Rhyzik (GR) 
\cite{gr}.  Note also GR 3.761.2, 3.761.4, and 4.422.)  Eq. (\ref{gr1}) 
implies that any constant term in the matrix element will give no contribution 
to the potential at $r\neq 0$.  Such constant terms do, however, lead to 
a $\dl$ contribution, as one sees by returning to the form (\ref{veq}).  
These delta-function contributions can 
be viewed as renormalizing the bare coupling strength.  (The $s$-channel 
diagram also contributes solely to the coupling-constant renormalization.)  
Another consequence of (\ref{gr1}) is that no dependence on $\Lambda$ 
survives in the $-1/r^3$ potential.  The result of substituting (\ref{mtexch}) 
into (\ref{veq2}) is thus:
\BE
\label{pot}
V_{t-{\rm exch}}(r) = - \frac{\lambda^2}{256 \pi^3 E^2} \frac{1}{r^3} .
\EE

    Allowing for the mass of the exchanged particles leads to \cite{ramond}
\BE
{\cal M}_{t-{\rm exch}}(q) = \frac{\lambda^2}{32 \pi^2} \left[ 
f(q) \ln \left( \frac{f(q) + 1}{f(q) - 1} \right) + {\rm const.} \right] ,
\EE
where $f(q) \equiv \sqrt{1+4m^2/q^2}$ and the constant term contains a 
$\ln \Lambda^2$ divergence, as in (\ref{mtexch}).  Substituting in 
(\ref{veq2}) we obtain the result in (\ref{pot}) multiplied by 
\BE
 I \equiv  - \frac{1}{\pi} \int_0^\infty dy \sin y \, \sqrt{y^2 + \sigma^2} \, 
\ln \left( \frac{ \sqrt{y^2 + \sigma^2} +y}{\sqrt{y^2 + \sigma^2} - y} \right) ,
\EE
where $\sigma = 2 m r$.  Thanks to (\ref{gr1}) this integral satisfies 
\BE
\frac{\partial I }{\partial \sigma} = -\frac{\sigma}{\pi} \int_0^\infty dy 
\frac{\sin y}{\sqrt{y^2 + \sigma^2} } \ln \left( 
\frac{ \sqrt{y^2 + \sigma^2} +y}{\sqrt{y^2 + \sigma^2} - y} \right) ,
\EE
which can be evaluated by using GR 3.775.1 and noting that 
$X^\nu - Y^\nu \to \nu \ln(X/Y)$ as $\nu \to 0$.  The result is 
$- \sigma K_0(\sigma)$, where $K_0$ is the modified Bessel function of 
order zero.  Integrating with respect to $\sigma$ gives 
\BE
I = \sigma K_1(\sigma).  
\EE
(We used formula 9.6.26 of Ref. \cite{as}.  The constant of integration 
must be zero to agree with the massless case ($\sigma =0$) evaluated earlier.)  
Therefore the massive case yields 
Eq. (\ref{pot}) times a factor $2mr K_1(2mr)$.  This factor becomes unity 
when $mr \to 0$, while for large $mr$ it tends to 
$\sqrt{\pi m r} \, {\rm e}^{-2 m r}$.  The exponential factor makes good sense; 
it is just like the Yukawa factor except that, as there are two particles 
exchanged, we get ${\rm e}^{-2 m r}$ rather than ${\rm e}^{- m r}$.  

      [One can also evaluate $V_{t-{\rm exch}}$ from Feinberg {\it et al}'s 
formulas:  Their spectral function $\rho_2$ is given as 
$-I(t) \phi(s,t)/(4 \pi)$ (see their Eqs. (2.52)--(2.54)), where the 
$\phi$ factor here reduces to just $\lambda^2$, and $I(t)$ for equal 
masses becomes $\sqrt{t-4m^2}/(8 \sqrt{t})$.  Inserting this into their 
Eq. (2.28b) (times $m^2/E^2$ to convert their $U(r)$ to $V(r)$) gives 
\BE
    V_{t-{\rm exch}} = \frac{1}{16 \pi^2 E^2} \frac{1}{r} 
\int_{4m^2}^\infty dt \left(- \frac{\lambda^2}{32 \pi} \right) 
\frac{\sqrt{t-4m^2}}{\sqrt{t}} \, {\rm e}^{- \sqrt{t} r}.
\EE
Changing the integration variable to $z = r \sqrt{t}$ leads to 
\BE
    V_{t-{\rm exch}} = - \frac{\lambda^2}{256 \pi^3 E^2} \frac{1}{r^3} 
\int_{2mr}^\infty dz \sqrt{z^2 - 4m^2 r^2} \; {\rm e}^{- z}.  
\EE
The integral trivially gives unity for $m=0$ and is easily evaluated in 
general using GR 3.387.6 to give $2mr K_1(2mr)$, confirming our 
previous result.]

     In summary; for $m$ small the potential consists of a $\dl$ 
`hard core' plus an attractive $-1/r^3$ tail out to distances of order 
$1/(2m)$, beyond which the potential is exponentially suppressed.

\setcounter{equation}{0}
\section{The hard-sphere Bose gas}

     In this section we temporarily ignore the $-1/r^3$ part of the 
potential and consider a gas of `atoms' interacting via a short-range, 
repulsive potential.  We follow the hard-sphere-Bose-gas analysis given 
in the textbook of K.~Huang \cite{huang}. (See also \cite{aaa} and the 
original references \cite{lhy}.)  This analysis describes Bose-Einstein 
condensation in a dilute atomic gas; a phenomenon only recently observed 
in a {\it tour de force} of experimental technique \cite{traps}.  

     Consider a box of volume $\vol$ containing $\num$ identical bosons 
(`atoms').  A convenient variable is the number density, $n \equiv \num/\vol$.  
[Huang uses the volume per particle, $v = \vol/\num$, which is just $1/n$.]  
The atoms interact via an interatomic potential, assumed to be short range,
but which may have any shape.  The point is that, in quantum mechanics, 
low-energy scattering is insensitive to the shape of the potential and can 
be characterized by a single parameter, the scattering length $a$ --- 
which is just (minus) the low-energy limit of the scattering amplitude 
\cite{fntehs}.  

     The analysis assumes small $a$ in two senses:
\BE
\label{dil}
 {\mbox{\rm diluteness:}} \, \, \quad n a^3 \ll 1,
\EE
\BE
\label{low}
 {\mbox{\rm `low energy':}} \, \, \quad k a \ll 1,
\EE
where $k$ is the wavenumber (or momentum, since we set $\hbar$ to unity).  
The `low energy' assumption means that the scattering is essentially pure
$s$-wave.  

     An important technical simplification, explained in detail by Huang, 
is that, in this approximation, one may replace the short-range, repulsive 
potential by a $\delta$-function ``pseudopotential'' \cite{huang}:
\BE
\label{pseud}
V_{\rm pseud}(r) = \frac{4 \pi a}{m} \dl 
\left( \frac{\partial}{\partial r} \right) r,
\EE
where the differential operator $( \frac{\partial}{\partial r} ) r$ acts 
to ``weed out'' any $1/r$ singularity in the wavefunction to which it is 
applied.  The Hamiltonian of the system [see Huang (10.124) and (13.86)] 
is then:
\BE
\label{hh}
H = - \frac{1}{2 m} \sum_{j=1}^{\num} \nabla^2_j + 
\frac{4 \pi a }{m} \sum_{i<j} \delta^{(3)}({\bf r}_i - {\bf r}_j) 
\frac{\partial}{\partial r_{ij}}  r_{ij}.  
\EE
This pseudopotential Hamiltonian is intended to be used {\it in Born 
approximation} only.  This is important because, if treated as a real 
potential, a repulsive $\delta$-function, of whatever strength, produces 
{\it zero} scattering in 3+1 dimensions.  (See Appendix A).  The normalization 
of the $\delta$-function term is such that, in Born approximation, it gives 
a scattering amplitude $-a$, and thus mimics the full effect of the true 
potential.  

     In the (NR) quantized-field representation the Hamiltonian is 
\BA
H & = & - \frac{1}{2 m} \int \! d^3r \,
\psi^{\dagger}({\bf r}) \nabla^2 \psi({\bf r}) 
\nonumber \\
& & 
+ \frac{2 \pi a }{m} \int \! d^3r_1 \! \int \! d^3r_2 \,
\psi^{\dagger}({\bf r_1}) \psi^{\dagger}({\bf r_2}) 
\delta^{(3)}({\bf r}_1 - {\bf r}_2) \frac{\partial}{\partial r_{12}}  
[r_{12} \psi({\bf r_1}) \psi({\bf r_2}) ].
\label{hqfr}
\EA
Substituting 
\BE
\label{psi}
\psi({\bf r}) = \sum_{{\bf k}} a_{{\bf k}} 
\frac{ {\rm e}^{i {\bf k}.{\bf r}} }{\sqrt{\vol}}, \quad \quad 
\left[ a_{{\bf k}}, a^{\dagger}_{{\bf k}'} \right] = \delta_{{\bf kk}'}, 
\EE
where the operators, $a_{\bf k}^{\dagger}, a_{\bf k}$, create and 
annihilate atoms in the free-particle state of wavenumber ${\bf k}$, 
this becomes 
\BE
\label{h}
H = \frac{1}{2m} \sum_{{\bf k}} k^2 a^{\dagger}_{{\bf k}} a_{{\bf k}} 
+ \frac{2 \pi a}{m \vol} \sum_{{\bf p}, {\bf q}} 
a^{\dagger}_{{\bf p}} a^{\dagger}_{{\bf q}} \frac{\partial}{\partial r}  
\left[ r \sum_{{\bf k}} {\rm e}^{i {\bf k}.{\bf r}} 
a_{{\bf p}+{\bf k}} a_{{\bf q}-{\bf k}} \right]_{r=0} .
\EE
For the ground state of the system (and for any low-lying excited states) 
one expects almost all the atoms to be in the ${\bf k}=0$ state.  Hence, 
the occupation numbers should satisfy $n_{{\bf k}=0} \sim \num$ and 
$\sum_{{\bf k} \neq 0} n_{{\bf k}}/\num \ll 1$.  Thus, one can effectively 
make the replacement $a_0 = a_0^\dagger = \sqrt{\num}$.  Dropping terms 
relatively suppressed by $1/\num$ leads to an effective Hamiltonian 
\cite{huang,mis}: 
\BE
\label{heff1}
H_{\rm eff} = \num \frac{2 \pi n a}{m} + 
\frac{1}{2m} \sum_{{\bf k} \neq 0}{\! '} 
\left[ ( k^2 + 8 \pi n a) a^{\dagger}_{{\bf k}} a_{{\bf k}} 
+ 4 \pi n a (a^{\dagger}_{{\bf k}} a^{\dagger}_{-{\bf k}} + 
a_{{\bf k}} a_{-{\bf k}} ) \right].
\EE
[The prime on the summation has to do with the $(\partial/\partial r) r$ 
subtlety; see below.]  

     The next step is to diagonalize $H_{\rm eff}$ by a Bogoliubov 
transformation.  That is, one introduces new operators 
$b_{\bf k}^{\dagger}, b_{\bf k}$ that are suitable linear combinations of 
$a_{\bf k}^{\dagger}, a_{\bf k}$:
\BE
\label{atob}
a_{\bf k} = \frac{1}{\sqrt{1-\alpha_k^2}} 
\left( b_{\bf k} - \alpha_k b_{- {\bf k}}^{\dagger} \right).
\EE
The $b_{\bf k} b_{-{\bf k}}$ and 
$b_{\bf k}^{\dagger} b_{-{\bf k}}^{\dagger}$ terms in $H_{\rm eff}$ 
are eliminated by choosing
\BE
\label{alpx}
\alpha_k = 1+x^2-x\sqrt{x^2+2},  \quad \quad 
x^2 \equiv \frac{2 m \omega_0(k)}{8 \pi n a},
\EE
where $\omega_0(k)$ is the original atom spectrum, $k^2/(2m)$.  
This yields 
\BE
\label{heff}
H_{\rm eff} = E_0 + 
\sum_{{\bf k} \neq 0} \omega(k) b_{\bf k}^{\dagger} b_{\bf k} ,
\EE
where 
\BE
\omega(k) = \omega_0(k) \left( \frac{1+\alpha_k}{1-\alpha_k} \right) 
= \omega_0(k) \sqrt{1 + 2/x^2},
\EE
which gives 
\BE
\label{bog}
\omega(k) = \omega_0(k) \sqrt{1 + (16 \pi n a)/(2 m \omega_0(k))}.  
\EE
The new operators, $b_{\bf k}^{\dagger}, b_{\bf k}$, create and destroy the 
elementary excitations of the Bose-Einstein condensate  --- phonons --- 
that are not single-atom excitations, but coherent states involving an 
indefinite number of atoms.  The spectrum of these excitations (their 
kinetic energy as a function of their momentum) is given by $\omega(k)$. 

     The ground-state energy is given by 
\BE 
\label{e0nr}
E_0 = \num \frac{2 \pi n a }{m} - 
\frac{2 \pi n a}{m} \sum_{{\bf k} \neq 0}{\! '} \,  \alpha_{k} 
= \num \frac{2 \pi n a }{m} \left( 1 + {\cal O}( \sqrt{n a^3} ) \right).
\EE
The first term arises simply from the ${\bf p} = {\bf q} = {\bf k} =0$ term 
in the summation in (\ref{h}).  The second term arises in the diagonalization 
step when one makes the substitution 
$b_{\bf k} b_{\bf k}^{\dagger} = b_{\bf k}^{\dagger} b_{\bf k} + 1$.  
The $\bf k$-summation can be converted to an integral by 
$\sum_{\bf k} \to \vol \! \int \! d^3 k/(2 \pi)^3$.  The integral would be 
linearly divergent but for the prime on the summation, whose effect is to 
subtract the divergent term \cite{fntepr}.  The result is that this second 
term is suppressed relative to the first by a factor of $\sqrt{na^3}$.

     An important observation \cite{csz} is that in the limit $a \to 0$, 
with $n \to \infty$ such that $n a$ remains finite, the results remain 
non-trivial while the approximations (\ref{dil}, \ref{low}) become 
{\it exact}:  The gas is then infinitely dilute, since 
$n a^3 = (n a) a^2 \to 0$, and the assumption $k \ll 1/a$ becomes no 
restriction.  In this limit the atoms' interaction vanishes, but 
non-trivial effects survive because there is an infinite density of them.  
One effect is the non-zero ground-state energy term in (\ref{heff}).  
The other is that the system's excitations are phonons, non-trivial 
coherent states of indefinitely many atoms, whose spectrum (\ref{bog}) 
is not the trivial $k^2/2m$ atom spectrum.  The phonons themselves are 
exactly non-interacting, so the resulting theory is ``trivial'' in the 
technical sense.  The relevance of this ``trivial''-but-not-entirely-trivial 
limit will become clear later.  

     Now let us consider the effect of allowing for relativistic kinematics.  
For actual atoms (Hydrogen, Lithium, etc.) this would be pointless, 
because their mass $m$ is always much greater than $1/a$; consequently, 
since the analysis assumes $k \ll 1/a$, one necessarily has $k$ much, much 
less than $m$.  However, we shall want to consider ``atoms'' whose mass is 
small compared to $1/a$.   Thus, even when the energy is `low' in the 
sense of $k \ll 1/a$ it need not be small with respect to $m$.  Thus, 
for us, relativistic kinematics will be relevant.  (We shall speak of 
`low energies' if $k \ll 1/a$ and `NR energies' if $k \ll m$.)  

     A quick-and-dirty argument for how to ``relativize'' Huang's results 
is the following.  (In Appendix D we shall describe a proper relativistic 
version of the calculation.)  
The original spectrum $\omega_0(k)$, instead of being the NR kinetic 
energy $k^2/2m$ should be the relativistic kinetic energy 
$E_k - m$, where $E_k \equiv \sqrt{{\bf k}^2 + m^2}$.  
(Recall that the relativistic and NR conventions for the zero of kinetic 
energy differ by $m$.)  As seen in the previous section, the formula for 
the scattering amplitude in terms of $V({\bf r})$, (\ref{fthv}), has a 
factor of $E$, while $V({\bf r})$ in (\ref{veq}) has a $1/(4E^2)$ factor.  
Therefore, with respect to the NR case, we must scale $a$ by a factor of 
$(E/m)(m/E)^2 = m/E$.  Hence, $x^2$ in Eq. (\ref{alpx}) becomes 
$2 E_k (E_k - m)/(8 \pi n a)$, and Eq. (\ref{bog}) becomes 
\BE
\tilde{E}_k  = (E_k - m) \sqrt{1 + \frac{8 \pi n a}{E_k (E_k-m)}}.
\EE
In the limit $m \to 0$, where the atom spectrum is $E_k=k$, the phonon 
spectrum $\tilde{E}_k$ tends to $\sqrt{k^2 + 8 \pi n a}$ \cite{cian}, which 
has the usual form for a relativistic particle.  We may then identify 
\BE
M^2(n) \equiv 8 \pi n a 
\EE
with the mass squared of the `phonon' excitation (which is the ``Higgs 
particle'' in QFT language).

\setcounter{equation}{0}
\section{Effect of the $-1/r^3$ potential}

     We now turn our attention to the effect of an attractive $-1/r^3$ 
potential between the atoms.  The special property of a $-1/r^3$ potential 
is actually evident at the classical level and it is instructive to look 
at this first.  Conveniently, the text by Reif \cite{reif} discusses the 
dilute classical gas using an illustrative interatomic potential of the form: 
\BE
V(r) = \left\{ \begin{array}{ll} \infty, \quad \quad \quad  & r < r_0 \\
                             -{\cal A}/r^s,                & r > r_0
               \end{array}
       \right.
\EE
with an arbitrary power $s$.  (${\cal A}$ is a constant, which Reif writes 
as $u_0 r_0^s$.)  The internal energy of the gas receives a 
contribution from pairwise interactions that is the number of pairs times 
the average energy of a pair: 
$\half \num(\num-1) \bar{u} \sim \half \num^2 \bar{u}$.  
Because the gas is dilute, 3-body interactions, etc., are negligible, 
and also the motion of any pair should not be appreciably correlated 
with the motion of the other atoms, which simply provide a heat bath of 
temperature $T$.  Then, the probability of a given separation $r$ is 
proportional to the Boltzmann factor, yielding \cite{reif}
\BE
\label{ubar}
\bar{u} = \frac{ \int \! d^3 r \, {\rm e}^{- \beta V(r)} \,V(r) }
               { \int \! d^3 r \, {\rm e}^{- \beta V(r)}},
\EE
where $\beta = 1/(k_B T)$, where $k_B$ is Boltzmann's constant.

      In the high-temperature (small $\beta$) limit one might naively 
say that the exponentials go to unity, giving 
\BE
\label{ubarht}
\bar{u} = \frac{1}{\vol} \int \! d^3 r \, V(r) = 
- \frac{4 \pi {\cal A}}{\vol} \int_{r_0}^{\infty} \frac{dr}{r^{s-2}} \, ,
\EE
which produces a logarithmic divergence in our case, $s=3$.  A more 
satisfactory derivation of this result \cite{reif} is to re-express 
(\ref{ubar}) as 
\BE
\bar{u} = - \frac{ \partial}{\partial \beta} 
\ln \left( \int \! d^3 r \, {\rm e}^{- \beta V(r)} \right).
\EE
Adding and subtracting unity from the integrand, one has
\BE
\int \! d^3 r \, {\rm e}^{- \beta V(r)} = \vol + I(\beta),
\EE
with 
\BE
I(\beta) \equiv \int \! d^3 r \left( {\rm e}^{- \beta V(r)} - 1 \right),
\EE
so that, up to ${\cal O}(1/\vol^2)$ terms, 
\BE 
\label{ui}
\bar{u} = - \frac{1}{\vol} \frac{ \partial I}{\partial \beta}.
\EE
The integral $I(\beta)$ is related directly, by $B_2 = - \half I(\beta)$, 
to the second virial coefficient, which represents the first deviation from 
the ideal-gas law in an expansion of the pressure in powers of the 
density:
\BE
\frac{p}{k_B T} = n + B_2 n^2 + {\cal O}(n^3).  
\EE
At high temperatures $I(\beta)$ can be evaluated by expanding the exponential, 
yielding 
\BE
I(\beta) = - \frac{4 \pi}{3} r_0^3 + 
4 \pi {\cal A} \beta \int_{r_0}^{\infty} \frac{dr}{r^{s-2}}.
\EE
The first term is the excluded volume due to the hard core.  Substituting 
in (\ref{ui}) reproduces the naive result (\ref{ubarht}), and we see again 
the divergence as $s \to 3$.  

     Now, it might seem odd to focus on the high-temperature limit, when 
we are really interested in $T=0$.  However, the divergence at $s=3$ will 
be present at any temperature; it is just easiest to see it in the 
high-temperature limit.  [For any $\beta$ the exponent $\beta V(r)$ becomes 
small as $r \to \infty$, allowing us to expand the exponential to obtain 
the $1/r^{s-2}$ behaviour at large $r$.]  Also, the high-temperature limit 
in the classical analysis has something in common with the $T=0$ quantum 
case.  If we consider a potential $V(r)$ quantum mechanically as a 
perturbation we obtain a contribution to the energy:
\BE 
\frac{1}{2} \int \! d^3r_1 \! \int \! d^3r_2 \, 
\psi^{\dagger}({\bf r_1}) \psi^{\dagger}({\bf r_2}) 
V(|{\bf r}_1 - {\bf r}_2|) \psi({\bf r_1}) \psi({\bf r_2}).
\EE
This is in the NR quantized field representation [Cf. Eq. (\ref{hqfr}), 
but with a potential $V(r)$ in place of the $\delta$-function 
pseudopotential].  For the zero-temperature ground 
state of the unperturbed gas we have simply $\psi({\bf r}) =$~const. 
$= \sqrt{\num/\vol}$ (normalized to 
$\int \! d^3r \, \psi^{\dagger} \psi =\num$).  
Thus, the above equation gives
\BE
\frac{1}{2} \left( \frac{\num}{\vol} \right)^2 \vol \int \! d^3 r V(r),
\EE
which is of the form $\half \num^2 \bar{u}$ with 
\BE
\bar{u} = \frac{1}{\vol} \int \! d^3 r V(r),
\EE 
just as in (\ref{ubarht}).  The point is that in the unperturbed quantum 
ground state the particles are evenly distributed over the box, just as 
they are in the classical, high-temperature case.  So in both cases we 
get an unweighted integral over $V(r)$.  This produces the logarithmic 
divergence when $V(r) \propto -1/r^3$.  

     The crucial question now is: ``What cuts off this divergence?''  
Clearly, the atom mass $m$ can act as a cutoff, since the actual potential 
only behaves as $-1/r^3$ out to distances of order $1/(2 m)$, due to 
the $2mr K_1(2mr)$ factor.  However, when $m$ becomes very small, and this 
cutoff distance becomes enormous, another effect comes into play; namely, 
the possibility of other atoms ``getting in the way'' of the interaction 
and tending to ``screen'' it.  The crucial point is that this provides a 
cutoff that involves the background density $n$.

     A first, over-simplistic approach, which nonetheless brings out the 
basic point that $r_{\rm max}$ will depend on $n$, is to suppose that the 
interaction becomes ``screened'' as soon as it is more likely than not 
that a third particle intervenes between the two particles that are trying 
to interact.  That would give an $r_{\rm max}$ of order the average atom 
spacing, which is $n^{-1/3}$.  

     A better argument is to say that in the ``fish'' diagram the virtual 
particles being exchanged have to propagate through the background medium.  
In so doing they will experience multiple collisions with the ${\bf k}=0$ 
background atoms.  This produces a propagator $G'$ that is a sum of all 
possible numbers of ``mass insertions,'' where each insertion 
$-i \Delta^2$ corresponds to a collision with a background atom:
\BE
G' = G + G(-i \Delta^2)G + G(-i \Delta^2)G(-i \Delta^2)G + \ldots \, ,
\EE
where $G=\frac{i}{p^2 - m^2}$, giving $G'= \frac{i}{p^2 - (m^2 + \Delta^2)}$.  
The new propagator must be the {\it phonon} propagator, since these are the 
elementary excitations in the presence of the background density, $n$, and are 
the only things that propagate over macroscopic distances.  Thus, we 
can identify $m^2 + \Delta^2$ as the phonon mass-squared $M^2(n)$.  
This implies a cutoff distance $r_{\rm max} \sim 1/(2 M(n))$, rather than 
$1/(2 m)$.  This assumes that $M(n)$ is much greater than $m$; otherwise the 
atom-mass cutoff $1/(2m)$ would take precedence.  Recalling that 
$M^2(n) \sim 8 \pi n a$ for $M(n) \gg m$, we see that 
$r_{\rm max} \propto n^{-1/2} a^{-1/2}$, rather than the $n^{-1/3}$ of the 
previous paragraph.

\setcounter{equation}{0}
\section{The ground-state energy density}

     We are now in a position to make an estimate of the ground-state 
energy density of the gas as a function of the atom density $n$.  
From the last section, the contribution from the $-{\cal A}/r^3$ 
interaction is $\half \num^2 \bar{u}$, with 
\BE
\bar{u} = \frac{1}{\vol} \int_{r_0}^{r_{\rm max}} 
\left( - \frac{{\cal A}}{r^3} \right) 4 \pi r^2 dr
= - \frac{4 \pi {\cal A}}{\vol} \ln(r_{\rm max}/r_0),
\EE
The coefficient of the $-1/r^3$ interaction, ${\cal A}$, can be inferred 
from Eq. (\ref{pot}).  Including a factor of 2 to allow for both $t$ and $u$ 
amplitudes, we have ${\cal A}=\lambda^2/(128 \pi^3 E^2)$.  (For the ground 
state, with almost all of the particles in the ${\bf k}=0$ mode, we may 
set $E=m$.)  Setting $r_{\rm max}$ equal to 
$1/(2 M(n)) = 1/(2 \sqrt{8 \pi n a})$ we have an energy contribution 
\BE 
\frac{1}{2} \num^2 
\left( - \frac{4 \pi}{\vol} \frac{\lambda^2}{128 \pi^3 m^2} \right) 
\frac{1}{2} \ln \left( \frac{1}{32 \pi n a r_0^2} \right) .
\EE

     In addition, there is the ground-state energy found in the hard-sphere 
Bose-gas analysis, namely $\num \frac{2 \pi n a}{m}$.  (See Eq. (\ref{e0nr}).)  
This represents the energy cost of the hard-sphere repulsions in the quantum 
ground state.  Finally, there is the rest-mass cost of having $\num$ particles 
(which, of course, is ignored in non-relativistic calculations).  In total 
we have 
\BE 
E_{\rm g.s.} = \num m + \num \frac{2 \pi n a}{m} + 
\num n \frac{\lambda^2}{128 \pi^2 m^2} \ln (32 \pi n a r_0^2 ).  
\EE
Dividing by the volume yields the energy density:
\BE
\label{enden}
{\cal E}_{\rm g.s.} = n m + \frac{2 \pi n^2 a}{m} + 
n^2 \frac{\lambda^2}{128 \pi^2 m^2} \ln (32 \pi n a r_0^2 ).  
\EE
We see that this is a sum of $n$, $n^2$ and $n^2 \ln n$ terms; these 
represent 1-body, 2-body, and ``2-body-plus-medium'' effects, respectively.  
Possible 3-body ($n^3$) or higher terms are negligible because of the 
diluteness assumption.  

     Note that if the attractive potential had fallen off faster than 
$1/r^3$ it would have given a simple $n^2$ term --- which could then have 
been combined with the other $n^2$ term.  That would correspond to replacing 
$a$, the scattering length due to the hard-core alone, by the scattering 
length produced by the whole potential.  This observation suggests an 
alternative way to obtain ${\cal E}_{\rm g.s.}$: simply replace the ``$a$'' 
in the hard-sphere term $2 \pi n^2 a/m$ by the ``effective scattering length'' 
$a_{\rm eff}(r_{\rm max})$ of the whole potential (hard core plus $-1/r^3$), 
cutting off the infrared divergence at $r_{\rm max}= 1/2 M(n)$.  
In Appendix B we show that $a_{\rm eff}$ can be expressed as 
$a - m {\cal A} \ln(r_{\rm max}/r_0)$, with the parameter $a$ now representing 
the scattering length due to the core {\it and} the ``short-range part'' of 
the $-1/r^3$ potential.  (Thus, $a$ can naturally be much greater than $r_0$).  
The $-m {\cal A} \ln(r_{\rm max})$ term in $a_{\rm eff}$ agrees precisely 
with the Born-approximation result --- not surprisingly, since it arises 
from the ``long-range tail'' of the $-1/r^3$ potential, where the potential 
is very weak.  Thus, our treatment of the $-{\cal A}/r^3$ interaction using 
lowest-order perturbation theory in Sect. 5 is effectively exact, provided 
that we re-interpret $a$.  

     Because of the $n^2 \ln n$ term in ${\cal E}_{\rm g.s.}$, the system has 
a first-order phase transition.  We shall analyze this in Sect. 8, but first 
we re-cast the result in QFT language.

\setcounter{equation}{0}
\section{Translation to QFT language}

     What do the scattering length $a$ and the density $n$ correspond to 
in QFT terms?  Clearly, the scattering length and the coupling $\lambda$ 
are related, since both reflect the strength of the interaction.  To leading 
order, we may directly compare the QFT expression for the cross section, 
(\ref{dsdoqft}), with the QM result $d \sigma/d \Omega = a^2$.  Including 
a factor of 2 for identical-particle reasons \cite{fnte2}, we have 
\BE 
\frac{(2 \lambda)^2}{256 \pi^2 E^2} = a^2 ,
\EE
or 
\BE
\label{aeq}
a = \frac{\lambda}{8 \pi E}.  
\EE
In the ground state, where almost all atoms have ${\bf k}\!=\!0$, we may set 
$E\!=\!m$.  The above result also follows by identifying the $\dl$ potential 
obtained from the bare vertex diagram, Eq. (\ref{vbare}), with the $\dl$ 
pseudopotential of the hard-sphere analysis, (\ref{pseud}).  The same 
$a \leftrightarrow \lambda$ connection (in the context of an O(2) theory, 
where the numerical factor is different) has been noted in Ref. 
\cite{bernstein}.  

     The other relation we need is between the background density $n$ 
and the strength of the background field $\phi = \langle \Phi \rangle$.  
In a finite-volume box, where the modes are discrete, one may expand the 
field $\Phi({\bf x},t)$ as:
\BE
\label{phime}
\Phi({\bf x},t) = \sum_{\bf k} \frac{1}{\sqrt{2 \vol E_k}} 
\left[ a_{\bf k} {\rm e}^{i {\bf k}.{\bf x}} + 
a^{\dagger}_{\bf k} {\rm e}^{-i {\bf k}.{\bf x}} 
\right]
\EE
where $a_{\bf k}$ carries a time-dependent factor ${\rm e}^{-i E_k t}$, 
with $E_k=\sqrt{{\bf k}^2 + m^2}$ and 
\BE
[ a_{\bf k}, a^{\dagger}_{\bf k} ] = \delta_{{\bf k},{\bf k}'}.
\EE
The number operator is 
\BE
\label{nop}
\hat{\num} = \sum_{\bf k}  a^{\dagger}_{\bf k} a_{\bf k} .
\EE
Following Ref. \cite{huang}, one argues that in the condensate state 
almost all the atoms are in the ${\bf k}=0$ mode, so 
$\num= a^{\dagger}_0 a_0$, and one may effectively regard 
$a_0$ (or rather $a_0^{\rm NR} \equiv a_0 {\rm e}^{i m t}$) as the c-number 
$\sqrt{\num}$.  Then from (\ref{phime}) one obtains the expectation value 
\BE
\phi = \langle \Phi \rangle = \frac{1}{\sqrt{2 \vol m}} 2 a_0^{\rm NR} = 
\sqrt{\frac{2\num}{\vol m}},
\EE
and hence $n \equiv \num/\vol$ is given by 
\BE
\label{neq}
n = \half m \phi^2.
\EE
With this identification, the trick of setting 
$a_0 = a_0^\dagger = \sqrt{\num}$ 
is equivalent to shifting the field by a constant $\phi$.  

    We may now consider both $M^2$ and ${\cal E}$ to be functions of 
$\phi^2$, rather than $n$.  Using the translations (\ref{aeq}) and (\ref{neq}) 
gives
\BE
M^2 = 8 \pi n a = 8 \pi (\half m \phi^2) \left( \frac{\lambda}{8 \pi m} \right) 
= \half \lambda \phi^2.
\EE
Similarly, the energy density (\ref{enden}) becomes
gives
\BE
\label{eest}
{\cal E}_{\rm g.s.} = \frac{1}{2} m^2 \phi^2 + 
\frac{1}{16} \lambda \phi^4 + 
\frac{1}{2} \frac{\lambda^2}{256 \pi^2} \phi^4 
\ln \left( 2 \lambda \phi^2 r_0^2 \right).
\EE
The energy density, expressed as a function of the field's expectation 
value, is of course the {\it effective potential}, $V_{\rm eff}(\phi)$, 
in QFT language.  We obviously recognise the first term as the usual 
mass term.  Up to numerical factors \cite{fntenfac}, the second term is the 
$(\lambda/4!) \phi^4$ term of the classical potential and the last term 
is the zero-point energy of free-field fluctuations about the background 
field $\phi$ (the ``Trace-Log'' term in the functional evaluation of the 
effective potential {\it \`a la} Jackiw \cite{jackiw}).  Recall that $r_0$, 
our short-distance cutoff, is proportional to $1/\Lambda$, where $\Lambda$ 
is a momentum-space ultraviolet cutoff.

\setcounter{equation}{0}
\section{The phase transition region}

     We now study the conditions needed for the system to be at or near 
the phase transition.  (Away from this region the physics would be totally 
dull.)  We prefer to use the QFT variables and then translate back to $a$ 
and $n$.  Because we have been somewhat cavalier about numerical coefficients 
let us start from:  
\BE
\label{mheq}
M^2(\phi_B) = c_0 \half \lambda \phi_B^2,
\EE
and 
\BE
\label{veffeq}
V_{\rm eff} = \frac{1}{2} m^2 \phi_B^2 + c_1 \frac{\lambda}{4!} \phi_B^4 
+ c_2 \frac{\lambda^2}{256 \pi^2} \phi_B^4 
\ln (c_0 \half \lambda \phi_B^2/\Lambda^2),
\EE
where $c_0, c_1, c_2$ are numerical coefficients of order 1, whose specific 
values won't change the general picture.  (We have added a `$B$' 
subscript to $\phi$ to emphasize that it is the (constant part of the) 
{\it bare} field.  Field re-scaling is discussed in Appendix C.) 

      If $V_{\rm eff}$ has non-trivial minima at $\phi_B = \pm v_B$, 
then $v_B$ must be a solution of $dV_{\rm eff}/d \phi_B =0$, which gives 
the equation:
\BE
\label{vbeqa}
m^2 + c_1 \frac{\lambda}{6} v_B^2 + 
c_2 \frac{\lambda^2}{64 \pi^2} v_B^2 
\left( \ln (c_0 \half \lambda v_B^2/\Lambda^2) + \frac{1}{2} \right) =0.
\EE
Let $v_0$ denote the solution in the special case $m^2 = 0$; it is given by 
\BE
\label{v0eq}
v_0^2 = \frac{2 \Lambda^2}{c_0 \lambda} {\rm e}^{-1/2} 
\exp \left( - \frac{32 \pi^2}{3 \lambda} \frac{c_1}{c_2} \right).
\EE
The original equation (\ref{vbeqa}) can then be re-written as:
\BE
\label{vbeq}
f(v_B^2) \equiv - c_2 \frac{\lambda^2}{64 \pi^2} v_B^2 
\ln (v_B^2/v_0^2) = m^2.
\EE
A graph of $f(v_B^2)$ climbs steeply from zero, has a maximum, and thereafter 
decreases, becoming negative when $v_B^2 > v_0^2$.  Equating this to $m^2$ 
we see that: (i) If $m^2$ is positive and larger than the maximum value 
of $f$, there will be no solution.  This corresponds to $V_{\rm eff}$ 
having a single minimum at $\phi=0$.  (ii) If $m^2$ is positive, but not 
too large, there will be two solutions for $v_B^2$.  This corresponds to 
$V_{\rm eff}$ having a minimum at $\phi=0$, then a maximum (at the smaller 
$v_B$ root) and then a minimum (at the larger $v_B$ root).  (iii) If $m^2$ 
is negative there is a unique solution, with $v_B$ greater than $v_0$.  
Here the origin is a maximum of $V_{\rm eff}$ and $v_B$ is a minimum.  

    Case (i) is of little interest, having no SSB.  Case (iii) is interesting 
and perfectly acceptable in QFT terms, but with $m^2$ negative (tachyonic 
`atoms') our particle-gas description is inappropriate (or, at least, loses 
its intuitive appeal).  We therefore focus on case (ii), where $m^2$ is 
positive but smaller than the maximum value of $f$.  Since $f$ is maximum at 
$v_B^2 = {\rm e}^{-1} v_0^2$, we need
\BE
m^2 < c_2 \frac{ \lambda^2}{64 \pi^2} v_B^2
\EE
for a non-trivial minimum of $V_{\rm eff}$ to exist.  A stronger condition 
is needed for this minimum to be lower than the minimum at the origin; 
i.e., for $V_{\rm eff}(v_B) < 0$:
\BE
m^2 < c_2 \frac{ \lambda^2}{128 \pi^2} v_B^2.
\EE
Recalling Eq. (\ref{mheq}) and that $M_h \equiv M(\phi_B = v_B)$, we see 
that $m^2$ must be of order $\lambda M_h^2$ or less.  

    From (\ref{mheq}) at $\phi_B = v_B$, we see that $M_h^2$ is of order 
$\lambda v_B^2$, and $v_B^2$ (for the root corresponding to a minimum 
of $V_{\rm eff}$) lies between ${\rm e}^{-1} v_0^2$ and $v_0^2$.  Thus, 
from (\ref{v0eq}) we find that 
\BE
\label{mheqb}
M_h^2 = {\cal O} \left( \Lambda^2 
\exp\left( -\frac{32\pi^2}{3\lambda} \frac{c_1}{c_2} \right) \right).
\EE
We want to take the cutoff $\Lambda$ to infinity, but we want to keep 
$M_h^2$ finite, so that the phonons (Higgs bosons) correspond to 
finite-mass particles.  The only way to achieve these requirements is 
to have $\lambda$ tend to zero as $\Lambda \to \infty$.  In fact, 
just by re-arranging (\ref{mheqb}) we see that $\lambda$ needs to 
behave as:
\BE 
\frac{\lambda}{16 \pi^2} \sim \frac{2}{3}\frac{c_1}{c_2} 
\frac{1}{\ln(\Lambda^2/M_h^2)}.
\EE
Thus, $\lambda$ tends to zero as $1/\ln \Lambda$.  The result of the 
previous paragraph that $m^2 \sim \lambda M_h^2$ means that $m^2$ 
is much, much smaller than $M_h^2$.  Since we require $M_h^2$ to be finite, 
which defines our physical mass scale, we need $m^2$ to go to zero as 
$1/\ln \Lambda$.  We also see that $v_B^2$ must go to infinity, 
like $\ln \Lambda$ so that $\lambda v_B^2$, and hence $M_h^2$, is finite.

    This behaviour, namely 
\BE
\label{gb1}
\lambda = {\cal O}(1/\ln \Lambda), \quad \quad 
m^2 = {\cal O}(1/\ln \Lambda), \quad \quad v_B^2 = {\cal O}(\ln \Lambda), 
\EE
(for $\Lambda$ in units of $M_h$) 
is needed if the physics is to remain interesting as $\Lambda \to \infty$.  
If $\lambda$ were larger, then $M_h$ would go to infinity (in particular, 
$M_h= {\cal O}(\Lambda)$ for any finite $\lambda$).  If $m^2$ were larger, 
then the $\half m^2 \phi^2$ term would completely dominate $V_{\rm eff}$, 
and we would simply have a theory of non-interacting atoms.  The interesting 
region is where one is sufficiently close to the phase transition, 
where the symmetric vacuum and the $v_B$ vacuum are in close 
competition.  The unusual feature is that the elementary excitations of 
these vacua, atoms and phonons, respectively, have vastly different masses; 
to keep the phonon mass finite, we must let the atom mass be vanishingly 
small.  

    Using the translations $a = \lambda/(8 \pi m)$ and 
$n= \half m \phi_B^2$, we see that 
\BE
\label{gb2}
a = {\cal O}(1/\sqrt{\ln \Lambda}), \quad \quad 
n_v = {\cal O}(\sqrt{\ln \Lambda}), 
\EE
where $n_v$ is the atom density in the SSB vacuum $\phi_B = v_B$.  Thus, 
the scattering length $a$ tends to zero (though it is much greater than 
the core size $r_0 \sim 1/\Lambda$.)  This result reflects the ``triviality'' 
property of the $\LP$ theory \cite{triv,triv2}.  However, $n_v a$ remains 
finite, because of course $M_h^2$ was required to be finite.  
Thus, the situation mentioned earlier \cite{csz} is automatically 
realized; the `diluteness' and `low energy' approximations 
(\ref{dil}, \ref{low}) become exact 
($n_v a^3 \to 0$, and $k a \to 0$ for any finite momentum $k$).
The phonons are exactly non-interacting in this limit, so the theory 
is ``trivial'' in the technical sense.  However, the phonons are non-trivial 
coherent states of the atoms, so some interesting physics is still 
present.  

    Moreover, this interesting physics can show up `experimentally.'  
The energy density difference between the SSB vacuum and the $\phi=0$ 
vacuum is finite in the physical units defined by $M_h$.  Therefore, 
if we heat the walls of our box to some critical temperature, finite 
in units of $M_h$, we can expect to see a phase transition.  In the 
high-temperature phase there is no Bose condensate and the excitations 
are the massless atoms.

\setcounter{equation}{0}
\section{All-orders considerations}

      We began by considering the interparticle potential produced by 
the simplest diagrams: The tree-level vertex gave a $\dl$ hard-core 
interaction, while the 1-loop ``fish'' diagram gave an additional $-1/r^3$ 
interaction (times $2mr K_1(2mr)$ if the mass is retained).  What about 
higher-order diagrams?  What contributions to the interparticle potential 
do they give?  The answer is that in the region of interest, namely 
for a $\lambda$ that is ${\cal O}(1/\ln (\Lambda/M_h) )$, they basically 
give only contributions of the same type, namely $\dl$ and $-1/r^3$.  
Thus, while higher orders can change the coefficients of these terms, 
they cannot modify the basic {\it form} of the interparticle potential.  

      To see this, consider a general $n$-loop 4-point diagram, neglecting 
the mass $m$.  In general this will be a sum of leading, sub-leading, \ldots 
\, logarithms of $\Lambda/q$, where $q$ is the momentum transfer:
\BE
{\cal M}_n = \lambda^{n+1} \left[ A_n (\ln \Lambda/q)^n 
+ B_n (\ln \Lambda/q)^{n-1} + \ldots \right],
\EE
where $A_n, B_n, \ldots$ are some numerical coefficients.  It will be 
convenient here to use units with $M_h=1$, so that $\ln \Lambda/M_h$ 
may be abbreviated as $\ln \Lambda$, etc..  Substituting 
$\ln \Lambda/q = \ln \Lambda - \ln q$, and recalling that we 
are interested in the case where $\lambda$ is proportional to 
$1/\ln \Lambda$, we see that ${\cal M}$ consists of (i) $q$-independent 
terms with one or more $1/\ln \Lambda$ factors; (ii) $\ln q$ terms 
with two or more $1/\ln \Lambda$ factors; (iii) $\ln^2 q$ terms 
with three or more $1/\ln \Lambda$ factors; (iv) etc.  These must be 
inserted in the 3-dimensional Fourier-transform integral, (\ref{veq}) or 
(\ref{veq2}), to find their contributions to the interparticle potential, 
$V(r)$.  We consider each in turn. (i) The $q$-independent terms clearly 
produce $\dl$ contributions. (ii) The $\ln q$ terms will produce 
$-1/r^3$ contributions, exactly as in Sect. 3. (iii) The $\ln^2 q$ 
terms, when we set $q=y/r$, involve $\ln^2 y - 2 \ln y \ln r + \ln^2 r$; 
when inserted in the $y$ integration (\ref{veq2}) these give, respectively, 
a $1/r^3$ term, a $(\ln r)/r^3$ term, and zero (for $r \neq 0$).  Thus, the 
only new contribution to $V(r)$ is a $(\ln r)/r^3$ term, which however is 
suppressed by one more power of $1/\ln \Lambda$ than the $-1/r^3$ terms.  
(iv) Similarly, the $\ln^p q$ terms in ${\cal M}_n$ can generate 
$(\ln^{p-1} r)/r^3$ terms in $V(r)$, but these are suppressed by $p-1$ 
powers of $1/\ln \Lambda$, relative to the $-1/r^3$ terms.  

     Thus, in aggregate, there is no change in the ``$-1/r^3+\dl$'' form 
of the potential, except at very short distances, $r \sim 1/\Lambda$  
(or extremely large distances, $r \sim \Lambda/M_h^2$).  The modification at 
very small distances is not unexpected; it will smooth the join between the 
hard-core and the $-1/r^3$ pieces of $V(r)$.  (Any modification at extremely 
large distances, $r \sim \Lambda/M_h^2$, would be physically irrelevant since 
the interaction is effectively cut off at distances of order 
$r_{\rm max} \sim 1/(2 M_h)$.)  

     To summarize: higher orders give contributions comparable in size to 
the lowest-order terms, but, up to negligible corrections, those contributions 
maintain the basic ``$-1/r^3+\dl$'' form of the interparticle potential.  
Our previous analysis remains valid provided that we understand `$\lambda$' 
as an effective coupling strength incorporating these all-order effects.  
We need not worry about how to actually calculate this $\lambda$ in terms 
of the original bare coupling(s) \cite{fntephi6}.  This is exactly like 
the NR Bose gas, where the short-range potential, whatever its shape, can 
be parametrized by a single parameter $a$, the actual scattering length, 
and one does not concern oneself with calculating $a$ from the original 
potential.  Note, however, that we should not require the effective 
$\lambda$ to be a finite parameter; it and $a$ have to be infinitesimally 
small.  

     Finally, we note that higher orders should not upset the relation 
between the coefficient ${\cal A}$ of the long-range $-{\cal A}/r^3$ 
potential and $\lambda$.  The long-range potential arises physically from 
two short-range repulsive interactions linked by quasi-free propagation of 
two virtual particles.  If $\lambda$ represents the actual strength of the 
short-range interaction, then ${\cal A}$ is proportional to its square.

\setcounter{equation}{0}
\section{Conclusions and outlook}

     Since our opening section provides an outline of the whole picture, 
we give only the briefest of summaries here:  The attractive $-1/r^3$ 
interparticle potential leads to an $n^2 \ln n$ term in the energy density, 
and hence to a $\phi^4 \ln \phi^2$ term in the effective potential, and 
hence to a {\it first-order} phase transition, occurring {\it before} 
$m^2$ reaches zero.  

     The form of effective potential obtained here agrees with our 
previous work \cite{cs,csz} and is, of course, just what one would obtain 
in the one-loop approximation \cite{cw}.  However, our point is that this 
form is effectively {\it exact} (in the relevant case 
$\lambda = {\cal O}(1/\ln \Lambda)$, as $\Lambda \to \infty$).  This is 
{\it not} because higher-loop contributions are negligible: in fact, 
all terms in the loop expansion then have the same size.  However, as 
explained in the previous section, higher-loop contributions, with 
$\lambda = {\cal O}(1/\ln \Lambda)$, give contributions of the same 
{\it form} as at one loop.  

     Just as in the NR Bose-gas analysis, the result hinges, not on any 
perturbative assumption, but on the diluteness ($na^3 \ll 1$) and `low 
energy' ($k a \ll 1$) approximations \cite{huang,aaa}, both of which 
become exact in the continuum limit.  It might appear that our treatment of 
the $-1/r^3$ interaction's effect in Sect. 5 relied on perturbation theory:
However, as explained in Appendix B, the Born approximation is effectively 
exact because the crucial $\ln(r_{\rm max})$ behaviour arises from the tail 
of the potential, where it is very weak.  An alternative calculation that 
is directly a relativistic version of Ref. \cite{huang}'s Bose-gas analysis 
is presented in Appendix D.  

     The origin of the crucial $\phi^4 \ln \phi^2$ term can be viewed 
in at least two ways.  In this paper it was obtained as the effect of a 
$-1/r^3$ interaction --- a relativistic-quantum modification of the original 
interparticle potential.  In the usual QFT calculation, however, it arises 
from the zero-point energy of the fluctuation field $\Phi(x)-\phi$.  These 
two descriptions, in ``particle'' and ``field'' languages, respectively, 
are just two complementary ways of looking at the same physics.  There is 
a parallel with the Lamb shift in QED which can be viewed as the effect on 
the energy levels of a relativistic-quantum modification of the 
electron-nucleus potential \cite{bethe}.  However, it can also be viewed 
as arising from the interaction of the electron with the zero-point 
fluctuations of the electromagnetic field \cite{fnteqed}.  

     Both in our previous QFT calculation and here we find that the theory is 
``trivial,'' in agreement with a large body of evidence \cite{triv,triv2}.  
That is, we find that scattering amplitudes (for `atoms' {\it or} `phonons') 
vanish as the cutoff goes to infinity.  However, because the physics of the 
vacuum involves infinitely many particles interacting infinitesimally 
weakly, there is a finite effect on $V_{\rm eff}$.  In QFT terms this effect 
is just the zero-point energy of the free fluctuation field $\Phi(x)-\phi$.  
It makes perfect sense that the effective potential of a ``trivial'' 
theory should be just the classical potential plus the zero-point energy 
of the ``trivial'' fluctuations.  This predicted form of the effective 
potential has been confirmed to high precision by recent lattice data  
\cite{agodi,aaccc,edin}.  

      As so often, the difficulty is not in understanding new ideas, but 
in freeing ourselves from the tyranny of old ones.  The conventional view 
(see, eg., \cite{cw,peskin,izdr}) has it that higher-loop effects change the 
$\phi^4 \ln \phi$ term in the effective potential to something like 
$\phi^4/\mid \! \ln \phi \! \mid$.  There is then no SSB until $m^2$ goes 
negative, and the transition is second order.  This so-called 
``RG improvement'' arises from a re-summation of the geometric series of 
``leading logs'' to obtain a ``renormalized coupling constant'' that runs 
according to 
$\lambda_R(\phi) = \lambda_R(\mu)/[1- b \lambda_R(\mu) \ln(\phi/\mu) ]$, where 
$\mu$ is some finite renormalization scale and $b= 3/(16 \pi^2)$.  
Absorbing the 1 into the scale of the logarithm gives 
$\lambda_R(\phi) = 1/(b \ln(\tilde{\mu}/\phi))$, where $\tilde{\mu}$ 
is some finite mass scale. The effective potential in this approach is then 
basically $\lambda_R(\phi) \phi^4$.  However, the Landau-pole problem, due 
to the absence of asymptotic freedom in perturbation theory, renders this 
approach unavoidably inconsistent, as we discuss in detail in Ref. 
\cite{csmpla}.  Moreover, precise lattice data for the effective potential 
cannot be fitted by any version of the ``RG-improved'' formula 
\cite{agodi,aaccc,edin}.  

     In the particle-gas picture, if one attempted to follow the 
``RG-improvement'' program, one would apply leading-log re-summation to 
the 4-point function, thereby obtaining a scattering matrix element 
${\cal M}(q)$ with a Landau pole.  Its 3-dimensional Fourier transform 
would then not exist, and one could not obtain any sensible result for 
the interparticle potential, $V({\bf r})$.  

     The particle-gas language provides a clue to what is missing in 
current applications of the RG method to the $\lambda \Phi^4$ case.  
Those approaches assume that the main effect of higher-order radiative 
corrections is to make the coupling constant ``run''; i.e., the form of 
the action remains basically the same at all scales, just with a $\lambda$ 
coupling that evolves.  However, while some higher-order effects do indeed 
renormalize the strength of the bare interaction, there are other contributions 
--- starting with the $t,u$-channel ``fish'' diagram --- that produce 
{\it qualitatively different physics}, namely long-range attraction.  
This effect will appear in different guises in different formalisms, but 
its inclusion is crucial.  

\vspace*{2mm}
\begin{center}
{\bf Acknowledgements}
\end{center}

\vspace*{-1.5mm}

     The recent experimental achievement of Bose-Einstein condensation 
in a dilute atomic gas \cite{traps} was an important stimulus to this 
work.  One of us (P.M.S.) would like to thank Randy Hulet, Henk Stoof, and 
John Ralston for discussions.

This work was supported in part by the U.S. Department of Energy under
Grant No. DE-FG05-92ER40717.

\newpage

\renewcommand{\theequation}{\Alph{section}.\arabic{equation}}

\section*{Appendix A: \quad QM scattering from a short-range potential}

\setcounter{section}{1}
\setcounter{equation}{0}

      We briefly review quantum-mechanical scattering theory for a 
short-range repulsive potential, which is important background to our whole 
discussion.  The scattering of two equal-mass particles is equivalent to 
the scattering of a particle of reduced mass $\mu = \frac{m}{2}$ from a 
fixed potential \cite{fntedist}.  We consider a potential 
\BE
\label{sqpot}
V(r) = \left\{ \begin{array}{ll}
V_0  \quad \quad          & r < r_0, \\
0                         & r > r_0, 
\end{array} \right.
\EE
with a high `aspect ratio' $2 \mu V_0 r_0^2 > 1$ \cite{fntear}.  
Ultimately, we shall be interested in the limit $r_0 \to 0$ where the 
interaction becomes pointlike.  In this limit the potential becomes 
\BE
\label{vw}
V(r) \to W \dl, \quad \quad \quad 
W \equiv \frac{4}{3}\pi r_0^3 V_0.
\EE

      For `low energies' $k \ll 1/r_0$, the scattering is entirely 
$s$-wave.  The cross section is then given by a single term of the 
partial-wave expansion:
\BE
\frac{d \sigma}{d \Omega} = \left| f(\theta) \right|^2 = 
\frac{1}{k^2} \left| {\rm e}^{i \delta_0} \sin \delta_0 \right|^2 ,
\EE
where $\delta_0$ is the $s$-wave phase shift, defined by the asymptotic 
behaviour of the radial wavefunction:
\BE
\label{phsht}
\chi(r) \equiv r \psi({\bf r}) \propto \sin(k r + \delta_0) \quad \quad 
{\rm as} \, \, r \to \infty.
\EE
A low-energy expansion of $k \cot \delta_0$ is the famous ``effective-range 
formula:''
\BE
\label{erform}
k \cot \delta_0 = - \frac{1}{a} + \frac{1}{2} r_{\rm eff} k^2 + {\cal O}(k^4),
\EE
where $a$ is the {\it scattering length} and $r_{\rm eff}$ is the 
{\it effective range}.  We may neglect the $r_{\rm eff}$ term at 
`low energies' $k \ll 1/a$ \cite{fntear}.  At these low energies $\delta_0$ 
is small and negative, $\delta_0 = - k a$, giving a scattering amplitude 
$f(\theta) = -a$.  

     For the specific potential (\ref{sqpot}) the Schr\"odinger equation 
for energy $E=k^2/(2 \mu)$ yields 
\BE
\label{chisol}
\chi(r) = \left\{ \begin{array}{ll}
A \sinh \alpha r + B \cosh \alpha r  \quad \quad & r < r_0, \\
C \sin k r + D \cos k r                                    & r > r_0, 
\end{array} \right.
\EE
where $A,B,C,D$ are constants and
\BE
\alpha = \alpha(k) \equiv \left( 2 \mu (V_0 - E) \right)^{1/2} .
\EE
The condition that $\psi({\bf r})$ is regular at ${\bf r}=0$ requires 
$\chi(r=0)=0$, and hence fixes $B=0$.  Matching conditions at $r=r_0$ give
\BE
A \sinh \alpha r_0 = C \sin k r_0 + D \cos k r_0 
\EE
\BE
\alpha A \cosh \alpha r_0 = k \left( C \cos k r_0 - D \sin k r_0 \right).
\EE
From the ratio of these two equations we obtain
\BE
\frac{D}{C} =  
\frac{[k \cos k r_0 \sinh \alpha r_0 - \alpha \sin k r_0 \cosh \alpha r_0 ]}
{[k \sin k r_0 \sinh \alpha r_0 + \alpha \cos k r_0 \cosh \alpha r_0 ]} .
\EE
Comparing (\ref{phsht}) and (\ref{chisol}) we see that $\tan \delta_0 = D/C$.  
Hence, from (\ref{erform}) we have 
\BE
a =  \lim_{k \to 0} \left( -\frac{1}{k \cot \delta_0} \right) = 
\lim_{k \to 0} \left( - \frac{D}{k C} \right) .
\EE
Hence, we obtain
\BE
\label{ar0}
a = r_0 \left( 1 - \frac{\tanh \alpha r_0}{\alpha r_0} \right) ,
\EE
with $\alpha \equiv \alpha(0) = (2 \mu V_0)^{1/2}$.  
Since the function $1-\tanh x/x$ lies between 0 and 1, we see that 
$a \le r_0$.  No matter how large we make $V_0$, the strength of the 
potential, the scattering length can never exceed $r_0$.  Thus, one has 
``triviality'' --- zero scattering amplitude for all finite energies --- 
in the limit $r_0 \to 0$.  

     In particular, in the $\delta$-function limit, $V(r) \to W \dl$ one has 
$\alpha r_0 \propto \sqrt{W/r_0} \to \infty$ and so $a = r_0 \to 0$.  
The fact that a $\delta$-function potential produces zero scattering in 3 
(or more) dimensions reflects the ``triviality'' of (NR) $\LP$ theory 
\cite{beg}.   

     In Born approximation, however, one would get 
\BE
\label{aborn}
a = \frac{\mu}{2 \pi} \int \! d^3 r \, {\rm e}^{-i {\bf k}.{\bf r}} \, V(r) 
= \frac{\mu}{2 \pi} \frac{4}{3} \pi r_0^3 V_0 = \frac{\mu}{2 \pi} W
\EE
which is finite if $W$ is finite.  One may re-write the Born result as
\BE
a = \frac{1}{3} (\alpha r_0)^2 r_0 
\EE
and recognize that perturbation theory corresponds to an expansion of the 
result (\ref{ar0}) for small $\alpha r_0$.  However, a $\delta$-function 
potential actually entails a infinitely large $\alpha r_0$, as noted 
above.

\section*{Appendix B: \quad QM scattering from a $-1/r^3$ potential}

\setcounter{section}{2}
\setcounter{equation}{0}

     In this appendix we treat quantum-mechanical scattering from a 
$-1/r^3$-plus-hard-core potential.  We shall denote the scattering length 
for such a potential by $a_{\rm eff}(r_{\rm max})$, since it is necessary to 
cut off the potential at large distances to avoid an infrared 
divergence:
\BE
V(r) = \left\{ \begin{array}{ll}
V_0                        & r < r_0, \\
-{\cal A}/r^3 \quad \quad  & r_0 < r < r_{\rm max}, \\
0                          & r > r_{\rm max}, 
\end{array} \right.
\EE

     In the region $r_0 < r < r_{\rm max}$ the Schr\"odinger equation for 
zero energy ($\chi(r)=r \psi(r)$):  
\BE 
\frac{d^2 \chi}{d r^2} + \frac{2 \mu {\cal A}}{r^3} \chi = 0
\EE
can be solved exactly.  By defining
\BE 
z \equiv 2 \left( \frac{2 \mu {\cal A}}{r} \right)^{1/2}  
\EE
and 
\BE
\chi(r) = w(z)/z,
\EE
the equation is transformed into
\BE
z^2 \frac{d^2 w}{d z^2} + z \frac{d w}{d z} + (z^2 - 1) w = 0,
\EE
whose solutions are the Bessel functions $J_1(z)$ and $Y_1(z)$ \cite{as}.  

    The zero-energy solution for $\chi(r)$ in the three regions is then
\BE
\chi(r) = \left\{ \begin{array}{ll}
A \sinh \alpha r  \quad \quad     & r < r_0, \\
( F J_1(z) + G Y_1(z) )/z \quad   & r_0 < r < r_{\rm max}, \\
\tilde{C} r + D                   & r > r_{\rm max}, 
\end{array} \right.
\EE
where $A,F,G,\tilde{C},D$ are constants and 
$\alpha \equiv \sqrt{2 \mu V_0}$.  The solution at large $r$ is 
to be regarded as the $k \to 0$ limit of $C \sin kr + D \cos kr$, 
and so $\tilde{C}$ is identified with $k C$.  Recall from the previous 
appendix that the scattering length is the $k \to 0$ limit of 
$-D/(kC)$, which is $-D/\tilde{C}$.  This slight cheat allows to obtain 
the scattering length without needing to solve the Schr\"odinger equation 
for a general energy.  

     Matching the solutions at $r_{\rm max}$ gives 
\BE
\tilde{C} r_{\rm max} + D = \frac{1}{z_m}( F J_1(z_m) + G Y_1(z_m) ) 
\EE
\BE
\tilde{C} = \frac{1}{2 r_{\rm max}}( F J_2(z_m) + G Y_2(z_m) ) ,
\EE
where $z_m$ is $z$ at $r=r_{\rm max}$.  From the ratio of these equations 
we obtain $a_{\rm eff} =-D/\tilde{C}$ as 
\BE
a_{\rm eff} = r_{\rm max} \left[ 1 - \frac{2}{z_m}
\frac{( F J_1(z_m) + G Y_1(z_m) )}{( F J_2(z_m) + G Y_2(z_m) )} \right] .
\EE
For large $r_{\rm max}$ ($\gg 2 \mu {\cal A}$) we may expand the Bessel 
functions for $z$ small \cite{as}: 
\BA
J_1(z) = \half z + {\cal O}(z^2), \quad \quad & & 
Y_1(z) = \frac{1}{\pi} 
\left( - \frac{2}{z} + z[ \ln(z/2) + \gamma - \half] + {\cal O}(z^3) \right)
\nonumber \\
J_2(z) = {\cal O}(z^2), & & 
Y_2(z) = \frac{1}{\pi} 
\left( - \frac{4}{z^2} - 1 + {\cal O}(z^2 \ln z) \right) .
\EA
A cancellation of leading terms leaves 
\BE
\label{aeff1}
a_{\rm eff}= r_{\rm max} \frac{z_m^2}{4} 
\left( 2 \ln(z_m/2) + 2 \gamma + \pi \frac{F}{G} \right) .
\EE

    The ratio $F/G$ can be found by a similar matching at $r=r_0$:
\BE
\frac{F}{G} = \frac{ z_0 \tanh \alpha r_0 Y_2(z_0) - 2 \alpha r_0 Y_1(z_0) }
{2 \alpha r_0 J_1(z_0) - z_0 \tanh \alpha r_0 J_2(z_0) }.
\EE
For small $r_0$ ($r_0 \ll 2 \mu {\cal A}$) we may expand the Bessel functions 
for large $z$ to obtain
\BE
\frac{F}{G} = \frac{ \kappa + \tan (z_0 - \frac{3 \pi}{4} )}
{ \kappa \tan (z_0 - \frac{3 \pi}{4} ) - 1} ,
\EE
where $\kappa \equiv z_0 \tanh \alpha r_0/(2 \alpha r_0)$.  Note that we 
have not assumed that $\alpha r_0$ is small.  Because of the 
$\tan (z_0 - \frac{3 \pi}{4} )$ factors, with $z_0$ large, the result is 
sensitive to $z_0$ {\it modulo} $2 \pi$, and so, from the definition of 
$z$, depends on the precise ratio of $2 \mu {\cal A}$ to $r_0$.  Thus, the 
ratio $F/G$ can take any value, and depends sensitively on the precise 
details of the hard-core potential and how it joins on to the $-{\cal A}/r^3$ 
potential.  

     Substituting for $z_m$ we may write the result (\ref{aeff1}) as 
\BE 
a_{\rm eff} = a - 2 \mu {\cal A} \ln(r_{\rm max}/r_0),
\EE
where $a \equiv 2 \mu {\cal A} \left[ \ln ( 2 \mu {\cal A}/r_0 ) + 
2 \gamma + \pi F/G \right]$.  This can be compared with the result in 
the Born approximation which is 
\BE
a_{\rm eff} = \frac{\mu}{2 \pi} \int \! d^3 r \, V(r) = 
a_{\rm Born} - 2 \mu {\cal A} \ln(r_{\rm max}/r_0) ,
\EE
where $a_{\rm Born} \equiv \frac{2}{3} \mu V_0 r_0^3$ (see (\ref{aborn})).  
Thus, we may effectively use the Born-approximation result provided we 
replace $a_{\rm Born}$ by the free parameter $a$.  We may think of $a$ as 
the {\it actual} scattering length due to the core.  (More precisely, it 
is due to the core {\it and} the ``un-Born part'' of the $-1/r^3$ potential 
\cite{fntegb}.)  Just as in the pseudopotential approach of \cite{huang}, 
all the details of the short-range potential can be parametrized by the 
single parameter $a$. 

     The fact that the Born approximation yields the correct 
$\ln r_{\rm max}$ term is natural, since this term arises from the 
large-$r$ tail where the potential {\it is} weak.  

     In the $\LP$ context the $-{\cal A}/r^3$ potential arises from two 
hard-core interactions connected by particle-pair exchange.  It follows 
that ${\cal A}$ will be proportional to $a^2$.  In other words, if the 
short-range interaction is parametrized by a coupling 
$\lambda \equiv 8 \pi m a$, there will be a $-1/r^3$ interaction of order 
$\lambda^2$.  We can correctly treat the $-1/r^3$ interaction in Born 
approximation even though we cannot rely on Born approximation to relate 
$\lambda$ to the original bare $\lambda$.

\section*{Appendix C: \quad Field re-scaling in `particle-gas' language}

\setcounter{section}{3}
\setcounter{equation}{0}

     In the phase-transition region, our effective potential (\ref{veffeq}), 
written in terms of $\phi_B$ is an extremely flat function, since the 
$\phi_B=0$ vacuum and the $\phi_B=\pm v_B$ vacua are infinitely far apart 
($v_B^2 = {\cal O}(\ln \Lambda)$), but differ in energy density only 
by a finite amount.  It is therefore natural to want to re-define the 
scale of the horizontal axis; i.e., to define a ``renormalized'' or 
``re-scaled'' field $\phi_R$.  

     To do this we introduced the following procedure \cite{cs,csz,cspl}:   
First we decompose the full field $\Phi_B(x)$ into its zero- and 
finite-momentum pieces:
\BE
\Phi_B(x) = \phi_B + h(x),
\EE
where $\int \! d^4x \, h(x) =0$.  Then we re-scale the zero-momentum part 
of the field: 
\BE
\phi_B^2 = Z_{\phi} \phi_R^2.
\EE
The finite-momentum modes, however, are not re-scaled.  The point is that 
for a ``trivial'' theory, as here, scattering theory and the Lehmann 
spectral decomposition require the wavefunction renormalization constant 
$Z_h$ (in $h_B(x) = \sqrt{Z_h} h_R(x)$) to be unity.  However, there is no 
constraint on $Z_{\phi}$, since there is no scattering theory for 
a zero-momentum mode --- the incident particles never reach each other.  
For the purpose of rendering $V_{\rm eff}$ a finite function of $\phi_R$, 
any $Z_{\phi}$ of order $\ln \Lambda$ would do: What determines the 
absolute normalization is the requirement that 
\BE
\label{concon}
\left. \frac{ d^2 V_{\rm eff}}{d \phi_R^2} \right|_{\phi_R=v_R} = M_h^2.
\EE
This is a standard renormalization condition, but the context here is 
rather unconventional.  In this Appendix we clarify its meaning using the 
`particle-gas' language.  

     First, consider a slight perturbation of the symmetric vacuum state 
(``empty box'').  We add a very small density $n$ of atoms, each with 
zero 3-momentum.  The energy density is now:
\BE
{\cal E}(n) = 0 + n m + {\cal O}(n^2 \ln n),
\EE
where the first term is the energy of the unperturbed vacuum state (zero); 
the second term is the rest-mass cost of introducing $\num$ particles, divided 
by the volume; and the third term is negligible if we consider a sufficiently 
small density $n$.  Thus, we obviously have the relation 
\BE
\label{enm}
\left. \frac{ \partial {\cal E}}{\partial n} \right|_{n=0} = m.
\EE
The equivalent in field language follows from the translation 
\BE
\label{nphi}
n = \half m \phi_B^2
\EE
and is that
\BE 
{\cal E}(\phi_B) \equiv V_{\rm eff}(\phi_B) = 0 + \half m^2 \phi_B^2 + 
{\cal O}(\phi_B^4 \ln \phi_B^2),
\EE
so that 
\BE
\label{d2vb}
\left. \frac{ \partial^2 V_{\rm eff}}{\partial \phi_B^2} \right|_{\phi_B=0} 
= m^2.
\EE

     Now, let us consider a slight perturbation of the SSB vacuum (the box 
filled with a spontaneously-generated condensate).  Before we perturb it, 
this state has a density $n_v$ of atoms, where $n_v$ is a (local) minimum 
of ${\cal E}(n)$.  From (\ref{nphi}) we have the translation 
$n_v = \half m v_B^2$.  This vacuum state, though complicated in terms 
of atoms, is simple in terms of phonons:  By definition it is just the state 
with no phonons.  We now perturb it by adding a small density $n'$ of 
phonons, each with negligible kinetic energy.  The energy density of the 
perturbed state is then 
\BE
\label{enp}
{\cal E}(n') = {\cal E}(n'=0) + n' M_h + \ldots,
\EE
where the first term is the energy density of the unpertubed state; the second 
term is the rest-mass cost of the added phonons; and any other terms 
from phonon interactions are negligible if $n'$ is small enough.  Thus, 
paralleling (\ref{enm}) we have 
\BE
\label{phmass}
\left. \frac{ \partial {\cal E}}{\partial n'} \right|_{n'=0} = M_h.
\EE
It is now natural to define a ``phonon field'' whose constant part, $f$, 
is related to the phonon density $n'$ by
\BE
\label{npf}
n' \equiv \half M_h f^2,
\EE
in analogy to (\ref{nphi}).  The ``renormalized field'' $\phi_R$ is simply 
this $f$ plus a constant.  A constant must be added if we want to have 
$\phi_R$ proportional to $\phi_B$.  Since, by definition, $f=0$ when 
$\phi_B=v_B$, we need
\BE
\phi_R \equiv f +v_R
\EE
with 
\BE 
\frac{v_R}{v_B} = \frac{\phi_R}{\phi_B} \equiv \frac{1}{Z_{\phi}^{1/2}}.
\EE
Now we can eliminate $f$ in favour of $\phi_R$ and re-write (\ref{npf}) 
as
\BE
\label{np}
n' = \half M_h (\phi_R - v_R)^2.
\EE
Hence, (\ref{enp}) can be re-written in field language as 
\BE
{\cal E}(\phi_R) \equiv V_{\rm eff}(\phi_R) =  V_{\rm eff}(\phi_R=v_R) 
+ \half M_h^2 (\phi_R - v_R)^2 + \ldots .
\EE
The crucial condition, Eq. (\ref{concon}), follows directly from this.  
It is simply the field-language equivalent of ``phonon-language'' equation, 
(\ref{phmass}), and just says that the phonon mass is, self-consistently, 
$M_h$.  It is also, of course, the SSB-vacuum counterpart to Eq. (\ref{d2vb}) 
for the symmetric vacuum.  

     The moral of this story is that the constant field $\phi_R - v_R$ is 
related to phonon density $n'$ in the same fashion that $\phi_B$ is related 
to the atom density $n$.  Note that there is a duality under
\BE
\mbox{\rm atoms} \leftrightarrow \mbox {\rm phonons}, \quad \quad 
\phi_B \leftrightarrow \phi_R - v_R, \quad \quad 
n \leftrightarrow  n', \quad \quad 
Z_{\phi} \leftrightarrow  Z_{\phi}^{-1}.
\EE
Physically, this means that, we may use either `atom' or `phonon' 
degrees of freedom to describe the theory.  Small excitations about the 
`phonon vacuum' are easily described in terms of phonons, but are 
complicated in terms of atoms --- and {\it vice versa}.  Moreover, 
just as we can describe the `phonon vacuum' as a complicated coherent 
state of atoms, we may also view the `atom vacuum' (empty box) as a 
complicated coherent state of phonons.  

     One can obtain various formulas for $Z_{\phi}$ from the above 
considerations.  There is no way to express $Z_{\phi}$ just in terms 
of phonon properties of the phonon vacuum, nor in terms of atom 
properties of the atom vacuum.  $Z_{\phi}$ is about the connection 
between these two states, and requires some knowledge of the atom 
properties of the phonon vacuum, or, dually, the phonon properties of 
the atom vacuum.  For instance, in field language, one may write 
\BE
Z_{\phi}^{-1} = \frac{1}{M_h^2} \left. 
\frac{\partial^2 V_{\rm eff}}{\partial \phi_B^2} \right|_{\phi_B=v_B}.
\EE
Translated to particle language this gives
\BE
Z_{\phi}^{-1} = \frac{2 m n_v}{M_h^2} \left. 
\frac{\partial^2 {\cal E}}{\partial n^2} \right|_{n=n_v},
\EE
involving how the energy density varies around the phonon vacuum when we 
vary the atom density $n$.  (There is also a `dual' version of both these 
formulas.)  Note that the first derivative 
$\left. \frac{\partial {\cal E}}{\partial n} \right|_{n=n_v}$, vanishes, 
since ${\cal E}(n)$ has a minimum at $n=n_v$.  However, in terms of 
phonons, when we vary $n'$ about $n'=0$ (which corresponds to the state 
$n=n_v$), we get Eq. (\ref{phmass}).  It follows that we must have 
$\left. \frac{\partial n'}{\partial n} \right|_{n=n_v} =0$; i.e., around 
the phonon vacuum, a small variation in atom density causes no first-order 
change in phonon density (nor in the energy density).  One can express 
$Z_{\phi}$ in terms of the second derivative as:
\BE
Z_{\phi}^{-1} = \frac{2m n_v}{M_h} 
\left. \frac{\partial^2 n'}{\partial n^2} \right|_{n=n_v}.
\EE
(Again, there is a `dual' formula).  

     $Z_{\phi}$ can be written in yet more ways.  Substituting 
$(\phi_R - v_R)^2 = (\phi_B - v_B)^2/Z_{\phi}$ into (\ref{np}), and 
then using $\phi_B = \sqrt{2n/m}$ from (\ref{nphi}) yields 
\BE
Z_{\phi} = \frac{M_h}{m} \frac{(\sqrt{n} - \sqrt{n_v})^2}{n'}.
\EE
Evaluated at $n=n_v$ this gives $0/0$, but evaluated at $n=0$ it gives 
\BE
Z_{\phi} = \frac{M_h}{m} \frac{n_v}{n'|_{n=0}},
\EE
which is a `self-dual' formula involving the ratio (density of atoms in 
the phonon vacuum)/(density of phonons in the atom vacuum).  The 
correctness of the formula becomes obvious if we use the $n, \phi_B$ 
and $n', \phi_R$ translations, since it then reduces to just 
$Z_{\phi} = v_B^2/v_R^2$.

\section*{Appendix D: \quad A tidier calculation}

\setcounter{section}{4}
\setcounter{equation}{0}

     Finally, to bring together all the pieces of the picture, we 
describe a relativistic version of Huang's hard-sphere Bose-gas analysis.  
We start with the Hamiltonian of the relativistic field theory, 
Eq. (\ref{hamiltonian}).  It is convenient to subtract a term 
$\mu \hat{\num}$, where $\hat{\num}$ is the number operator 
$\sum a^{\dagger}_{\bf k} a_{\bf k}$.  In the end we shall set $\mu=0$.  
However, when comparing with the NR analysis we must set $\mu =m$ to 
convert to the NR convention that particle rest masses do not count in 
the kinetic energy --- so that in the total energy of the system, $H$, 
we must subtract $m$ for each particle in the system; i.e., 
$H_{\rm NR} = H - m \hat{\num}$.  

     The calculation then proceeds as in Sect. 4.  Instead of Eq. 
(\ref{psi}) for the NR field we have Eq. (\ref{phime}) for the relativistic 
field \cite{fntepsiphi}.  Making this substitution and writing 
$a_0 \sim a^\dagger_0 \sim \sqrt{\num}$ (which is equivalent to shifting the 
field by a constant, $\phi$) leads to  
\BE
\label{heffrel}
H - \mu \hat{\num} = \vol \left[ (m-\mu)n + \frac{\lambda n^2}{6 m^2} \right] 
+ \sum_{k \neq 0} \left[ 
a^\dagger_{\bf k} a_{\bf k} \left( E_k - \mu + \frac{\xi}{2 E_k} \right) + 
\frac{\xi}{4 E_k} \left( a_{\bf k} a_{-{\bf k}} + 
a^\dagger_{\bf k} a^\dagger_{-{\bf k}} \right) \right]
\EE
with 
\BE
\xi = \frac{\lambda n}{m} = \half \lambda \phi^2
\EE
Some comments on this are in order:  
Our $H$ was normal ordered (with respect to mass $m$), so the 
$\lambda$-independent terms are rather obvious.  The other terms come 
from $\int d^3x \, : \! \Phi^4 \! :$~.  Just as in the NR calculation, one 
needs to keep terms with at least two $a_0$ or $a_0^\dagger$ factors; other 
terms will be relatively suppressed by $1/\sqrt{\num}$ factors.  That is (in 
the words of Ref. \cite{aaa}) we need only take into account interaction 
of particles in the condensate with each other and interaction of 
``excited'' particles with particles in the condensate, neglecting 
interaction of ``excited'' particles with each other.  Finally, we observe 
that in the NR case, substituting $\mu = m$, and $E_k \sim m$, and 
$E_k - m \approx k^2/(2m)$ this equation reproduces Eq. (\ref{heff1}) 
(except for a different ground-state-energy term \cite{fntenf}) 
if we identify $\xi$ with $8 \pi n a$.  This implies the identification 
$\lambda = 8 \pi m a$, in agreement with (\ref{aeq}) above.

     We now proceed to introduce $b_{\bf k}, b^\dagger_{\bf k}$, just as in 
Eq. (\ref{atob}).  Requiring the $b_{\bf k} b_{-{\bf k}}$ terms to cancel 
gives 
\BE
\label{alpharel}
\alpha_k = 1+x^2-x\sqrt{x^2+2},  \quad \quad 
x^2 \equiv \frac{2}{\xi} E_k (E_k - \mu),
\EE
which has the same form as (\ref{alpx}), except for the modified definition 
of $x^2$ (which has the form anticipated in the final paragraph of Sect.4).  
We are then left (after a little algebra) with 
\BE
\label{hii}
H_{\rm eff} - \mu \hat{\num} = {\rm const.} + \frac{\xi}{4} \sum_{k \neq 0} 
\left[ b^\dagger_{\bf k} b_{\bf k} \frac{1}{E_k \alpha_k} 
- b_{\bf k} b^\dagger_{\bf k} \frac{\alpha_k}{E_k} \right],
\EE
and hence 
\BE
\label{heffrelb}
H_{\rm eff} - \mu \hat{\num} = E_0 + \sum_{k \neq 0} \tilde{E}(k)
b^\dagger_{\bf k} b_{\bf k},
\EE
which is the analog of (\ref{heff}).  The new spectrum $\tilde{E}_k$ is 
given by
\BE
\tilde{E}_k = (E_k - \mu) \sqrt{ 1 + \frac{\xi}{E_k(E_k - \mu)} }
\EE
For $\mu=m$ this gives the right NR limit.  With $\mu=0$ one 
gets 
\BE
\tilde{E}(k) = \sqrt{E_k^2 + \xi} = \sqrt{{\bf k}^2 + m^2 + \xi}
\EE 
which has the expected relativistic form.  We may then identify the 
`phonon' (`Higgs') mass squared as 
\BE
M^2(\phi) = m^2 + \xi = m^2 + \frac{1}{2} \lambda \phi^2.
\EE
This is the same as before, except that we have been more careful and 
retained the $m^2$ term.  However, this makes no real difference, since 
in the interesting region $m^2$ is infinitesimal in comparison with 
$\xi= \half \lambda \phi^2$.  

     The $E_0$ term in (\ref{heffrelb}) is
\BE
E_0 = 
\vol \left[ (m-\mu)n + \frac{\lambda n^2}{6 m^2} \right] - 
\frac{\xi}{4} \sum_{k \neq 0} \frac{\alpha_k}{E_k}, 
\EE
where the term in square brackets (for $\mu=0$) is just the classical 
potential, as one sees by  substituting $n=\half m \phi^2$.  The last term, 
which arises in the step from (\ref{hii}) to (\ref{heffrelb}), is the direct 
analog of the second term in (\ref{e0nr}) in the NR calculation.  The NR 
version has $m$ rather than $E_k$ in the denominator, and its ${\bf k}$ 
summation would be linearly divergent, but for the $(\partial/\partial r)~r$ 
pseudopotential subtlety.  The relativistic analog, because of the extra 
$1/E_k$ factor is only logarithmically divergent.  It represents a 
quantum contribution to the ground-state energy --- and it is not hard to 
recognize it as the effect we have discussed, arising physically from 
the quantum-induced $-1/r^3$ interaction between the atoms.  

     Rewriting the previous equation (for $\mu=0$), the energy density 
of the ground state as a function of $n$, or equivalently, $\phi$, is:
\BE
V_{\rm eff} = E_0/\vol = \frac{1}{2}m^2 \phi^2 + \frac{\lambda}{4!} \phi^4 
- \frac{1}{8} \lambda \phi^2 \int \! \frac{d^3 k}{(2 \pi)^3}
\frac{\alpha_k}{E_k}.
\EE
Substituting for $\alpha_k$ from (\ref{alpharel}), and noting that 
$x\sqrt{x^2 + 2} = 4 E_k \tilde{E}_k/(\lambda \phi^2)$ one has
\BE
\label{veehehe}
V_{\rm eff} = \frac{1}{2}m^2 \phi^2 + \frac{\lambda}{4!} \phi^4 
+ I_1(M) - I_1(m) - \frac{1}{4}\lambda \phi^2 I_0(m),
\EE
where
\BE 
I_1(M) \equiv \int \! \frac{d^3 k}{(2 \pi)^3} \frac{1}{2} \tilde{E}_k, 
\quad \quad 
I_1(m) \equiv \int \! \frac{d^3 k}{(2 \pi)^3} \frac{1}{2} E_k, 
\quad \quad 
I_0(m) \equiv \int \! \frac{d^3 k}{(2 \pi)^3} \frac{1}{2 E_k}. 
\EE
In field language $I_1(M)$ is just the zero-point energy contribution 
from a free field with a mass $M(\phi)$, while the last two terms in 
(\ref{veehehe}) are just the subtractions implied by normal ordering.  
These terms remove the quartic and quadratic divergences in $I_1(M)$, 
leaving a logarithmic divergence.  Explicitly, in the $m \to 0$ limit 
the result is 
\BE
I_1(M) - I_1(m) - \frac{1}{4}\lambda \phi^2 I_0(m) \to 
\frac{\lambda^2}{256 \pi^2} \phi^4 
\left[ \ln (\half \lambda \phi^2 /\Lambda^2 ) - \frac{3}{2} \right] .
\EE
Thus, the result for $V_{\rm eff}$ is our old friend the ``one-loop'' 
result.  

     {\it The crucial point is that this form of $V_{\rm eff}$ is effectively 
exact}.  Just as in the NR calculation the validity of the result 
requires only the `diluteness' ($na^3\to 0$) and `low-energy' 
($ka \to 0$) approximations.  The only subtlety is that the scattering 
length $a$ should 
be the {\it actual} scattering length.  The calculation of the actual $a$ 
in terms of the original potential may be very hard, but is of no 
importance.  In fact the actual shape of the original potential is irrelevant, 
(provided only that it is short-range and repulsive); its effect is 
entirely parametrized by the scattering length it gives rise to.  
Similarly, here the `$\lambda$' has to be understood as some effective coupling 
strength, incorporating the effects of the original QFT interaction 
(which could include $\Phi^6$, $\Phi^8$, \ldots, terms) to all orders.
In the QFT case, if a continuum limit is to exist (i.e., if finite physics 
is to exist in the limit $\Lambda \to \infty$), then $a$ and $\lambda$ must 
be infinitesimally small, which corresponds to ``triviality'' of the theory.  
However, there is still non-trivial physics for the ground state, due to 
an infinite number of infinitesimally weakly interacting particles.


\begin{thebibliography} {99}
\bibitem{fnteid}
Note that we are ignoring identical-particle complications here.
\bibitem{fnterel}
One may also understand the $E/m$ factor as follows:  In obtaining the 
Born-approximation cross section via Fermi's Golden Rule, the only changes 
are to the flux factor and to the density of states.  The flux factor, 
appearing in the denominator, is proportional to velocity, which is now 
$p/E$ instead of $p/m$.  In the free-particle density of states the 
counting of states in an interval $dp$ goes as usual, but when relating 
$dp$ to $dE$ we must use $E^2=p^2+m^2$ rather than the NR form 
$E-m = p^2/(2m)$, giving $p dp = E dE$ rather than $p dp = m dE$.  
We therefore obtain two factors $E/m$ in the cross section, and hence 
one factor of $E/m$ in the scattering amplitude.
\bibitem{fntehs}
In this sense, any short-range potential with scattering length $a$ is 
equivalent to a ``hard-sphere'' potential ($V(r) = \infty$ for $r<a$ and 
$V(r) = 0$ for $r>a$).  This corresponds to viewing the atoms as hard 
spheres with radius $a/2$, the closest approach distance being twice 
the radius.  However, in general $a$ depends on the strength of the 
potential as well as its spatial extent.  It will be important 
later not to confuse $a$ with the hard-core radius, $r_0$, of our 
$-1/r^3$-plus-hard-core potential.
\bibitem{mis}
The equation quoted is (13.94) of Huang \cite{huang}.  Beware some 
misprints in the preceding equation, (13.93).
\bibitem{fnte2}
Identical-particle factor-of-2 issues are a headache that we shall not 
attempt to resolve fully.  However, we argue that $\mid\!{\cal M}\!\mid^2$ 
in (\ref{dsdoqft}) should become 
$\mid\!{\cal M}(\theta) + {\cal M}(\pi-\theta)\!\mid^2 = (2 \lambda)^2$.  
The corresponding modification of the QM formula 
(\ref{dsdoqm}) seems to be already allowed for in Huang's $a$; that is, 
$\mid \! f(\theta) + f(\pi - \theta) \! \mid^2 \equiv a^2$.
\bibitem{fntephi6}
We could allow the original Lagrangian to include $\Phi^6$, $\Phi^8$, 
$\ldots$, terms, which would also modify the effective $\lambda$ but 
would not change the physics.
\bibitem{fntedist}
We shall again ignore indistinguishable-particle complications.
\bibitem{fntear}
The condition $2 \mu V_0 r_0^2 > 1$ means that for `low energies' 
$k \ll 1/r_0$ the particle's kinetic energy cannot exceed the barrier 
height $V_0$.  This precludes obtaining an effective range 
$r_{\rm eff} \gg r_0$.
\bibitem{fntegb}
Note that it is perfectly possible to have $a \gg r_0$.  In fact, one 
may check that the behaviour envisaged in Sect. 8, Eqs. (\ref{gb1}), 
(\ref{gb2}), in which $a \sim {\cal O}(1/\sqrt{\mid \! \ln r_0 \! \mid } )$, 
is consistent with the generic case that $F/G$ is an order-1 number.
\bibitem{fntenf}
This appears to bear out our comments on Eq. (\ref{eest}) in footnote 
\cite{fntenfac} concerning the role of the $s$-channel interaction.
\end{thebibliography}
\end{document}